\documentclass[twocolumn]{aastex631}

\def\smass{{$M_*$}}

\def\halpha{${\rm H}\alpha$}
\def\hbeta{${\rm H}\beta$}
\def\mbh{$M_{\rm BH}$}
	
\def\hst{{\it HST}}

\shorttitle{Comparing simulations and observations of black hole - galaxy relations}
\shortauthors{Ding et al.}
\graphicspath{{./}{figures/}}

\begin{document}

\title{Concordance between observations and simulations in the evolution of the mass relation between supermassive black holes and their host galaxies}

\author[0000-0001-8917-2148]{Xuheng Ding}
\affiliation{Kavli Institute for the Physics and Mathematics of the Universe, The University of Tokyo, Kashiwa, Japan 277-8583 (Kavli IPMU, WPI)}

\author[0000-0002-0000-6977]{John D. Silverman}
\affiliation{Kavli Institute for the Physics and Mathematics of the Universe, The University of Tokyo, Kashiwa, Japan 277-8583 (Kavli IPMU, WPI)}

\author[0000-0002-8460-0390]{Tommaso Treu}
\affiliation{Department of Physics and Astronomy, University of California, Los Angeles, CA, 90095-1547, USA}

\author[0000-0002-1605-915X]{Junyao Li}
\affiliation{CAS Key Laboratory for Research in Galaxies and Cosmology, Department of Astronomy, University of Science and Technology of China, Hefei 230026, China}
\affiliation{School of Astronomy and Space Science, University of Science and Technology of China, Hefei 230026, China}
\affiliation{Kavli Institute for the Physics and Mathematics of the Universe, The University of Tokyo, Kashiwa, Japan 277-8583 (Kavli IPMU, WPI)}

\author[0000-0002-7080-2864]{Aklant K. Bhowmick}
\affiliation{Dept. of Physics, University of Florida, Gainesville, FL 32611, USA}

\author[0000-0002-4096-2680]{Nicola Menci}
\affiliation{INAF Osservatorio Astronomico di Roma, via Frascati 33, I-00078 Monteporzio, Italy}

\author[0000-0002-3216-1322]{Marta Volonteri}
\affiliation{Institut d'Astrophysique de Paris, Sorbonne Universit\'e, CNRS, UMR 7095, 98 bis bd Arago, 75014 Paris, France}

\author[0000-0002-2183-1087]{Laura Blecha}
\affiliation{Dept. of Physics, University of Florida, Gainesville, FL 32611, USA}

\author[0000-0002-6462-5734]{Tiziana Di Matteo}
\affiliation{McWilliams Center for Cosmology, Dept. of Physics, Carnegie Mellon University, Pittsburgh, PA 15213, USA}

\author[0000-0003-0225-6387]{Yohan Dubois}
\affiliation{Institut d'Astrophysique de Paris, Sorbonne Universit\'e, CNRS, UMR 7095, 98 bis bd Arago, 75014 Paris, France}

\begin{abstract}
We carry out a comparative analysis of the relation between the mass of supermassive black holes (BHs) and the stellar mass of their host galaxies at $0.2<z<1.7$ using well-matched observations and multiple state-of-the-art simulations (e.g., Massive Black II, Horizon-AGN, Illustris, TNG and a semi-analytic model). The observed sample consists of 646 uniformly-selected SDSS quasars ($0.2 < z < 0.8$) and 32 broad-line active galactic nuclei (AGNs; $1.2<z<1.7$) with imaging from Hyper Suprime-Cam (HSC) for the former and Hubble Space Telescope (HST) for the latter. We first add realistic observational uncertainties to the simulation data and then construct a simulated sample in the same manner as the observations. Over the full redshift range, our analysis demonstrates that all simulations predict a level of intrinsic scatter of the scaling relations comparable to the observations which appear to agree with the dispersion of the local relation. Regarding the mean relation, Horizon-AGN and TNG are in closest agreement with the observations at low and high redshift ($z\sim$ 0.2 and 1.5, respectively) while the other simulations show subtle differences within the uncertainties. For insight into the physics involved, the scatter of the scaling relation, seen in the SAM, is reduced by a factor of two and closer to the observations after adopting a new feedback model that considers the geometry of the AGN outflow. The consistency in the dispersion with redshift in our analysis supports the importance of both quasar- and radio-mode feedback prescriptions in the simulations. Finally, we highlight the importance of increasing the sensitivity (e.g., using the James Webb Space Telescope), thereby pushing to lower masses and minimizing biases due to selection effects.
\end{abstract}
\keywords{Galaxy evolution (594); Active galaxies (17); Active galactic nuclei (16)}

\section{Introduction} \label{sec:intro}
The close correlations between the mass of supermassive black holes (BHs), \mbh, and the properties of their host galaxies (e.g., stellar mass, \smass) indicate a physical coupling during their joint evolution~\citep{Mag++98, F+M00, M+H03, H+R04, Gul++09}. To understand the nature of this connection, considerable efforts have been focused on measuring such correlations using broad-line active galactic nucleus (AGNs) over a range of redshifts, with the intention to determine how and when the correlation emerges and evolves over cosmic time. While an {\it observed} evolution has been found since redshift $z\sim2$ in galaxies with stellar mass \smass\ $\gtrsim 10^{10}M_{\odot}$ \citep[e.g.,][]{Tre++04, Peng2006a,Tre++07,Woo++08,Bennert11,Park15} in which the growth of BHs predates that of the host, other studies~\citep[e.g.,][]{Schramm2013, Sun2015, 2021ApJ...906..103L} predict that BHs grow commensurately with galaxies.
However, to understand the significance {\it intrinsic} evolution, it is necessary to take into account systematic uncertainties and the selection effects~\citep{Tre++07, Lauer2007, Schulze2014, Park15,Jahnke2009,  Ding2020,Li2021b}. 

Various theoretical models have been proposed to explain the origin of the scaling relations. For example, AGN feedback is considered as one of the possible viable mechanisms. During this process, a fraction of the AGN energy is injected into its surrounding gas, which can then regulate the mass growth of the BH and its host. In this scenario, star formation is inhibited by the heating and unbinding of a significant amount of gas. Alternatively, the mass relations can be explained through an indirect connection in which AGN accretion and star formation are fed through a common gas supply~\citep{Cen2015, Menci2016, angles_black_2017}. Actually, even without any physical mechanisms, statistical convergence from galaxy assembly alone (i.e., dry mergers) could instill the observed correlations~\citep{Peng2007, Jahnke2011, Hirschmann2010}. However, as expected from the central limit theorem, a higher dispersion would appear in the scaling relations at high-$z$ compared to what is observed today~\citep[e.g.,][]{Ginat2016, 2020MNRAS.498.5652K}. 

Numerical simulations provide an opportunity to further understand the connection between BHs and their host galaxies. For example, a comparison of scaling relations has been made using state-of-the-art cosmological hydrodynamical simulation of structure formation ({\tt MassiveBlackII}) and observational measurements at  $0.3<z<1$ \citep[e.g., ][]{DeG++15}, which show a positive evolution (i.e., the mass growth of the BH predates that of its host). Further efforts are using large-volume simulations to investigate the scaling relations and find good agreement with the local relation with redshift evolution, 
including the Magneticum Pathfinder smooth-particle hydrodynamics (SPH) Simulations~\citep{Steinborn2015}, the Evolution and Assembly of Galaxies and their Environments~(EAGLE) suite of SPH simulations~\citep{Schaye2015, Crain2015, McAlpine2016}, Illustris moving-mesh simulation~\citep{Genel2014, 2014MNRAS.444.1518V, Sijacki2015, Nelson2015, Li2019}, the Horizon-AGN simulation~\citep{2014MNRAS.444.1453D, 2016MNRAS.463.3948D, 2016MNRAS.460.2979V}, and the SIMBA simulation~\citep{Thomas2019, Dave2019}.
In particular, the \mbh-\smass\ relation using BH populations using six large-scale cosmological simulations (i.e, Illustris, TNG100, TNG300, Horizon-AGN, EAGLE, and SIMBA) has been compared with observations in the local universe in a recent study~\citep{Habouzit2021}. However, these comparison works are limited by the observation data in terms of the sample size ($<$100) and redshift range (i.e., limited to the local universe).

For such comparisons using simulations, it is crucial to consider the systematic uncertainties and selection biases. A direct means to account for these is to apply the same effects and selection to the simulation products and make a forward comparison in the observational plane. In~\citet{Ding2020b}, a direct comparison has been performed using 32 X-ray-selected AGN at $1.2<z<1.7$ and a direct comparison with two state-of-the-art simulation efforts, including {\tt MassiveBlackII} (MBII) and a Semi-analytic Model \citep[SAM,][]{Menci2014, Menci2016}. The dispersion in the mass ratio between BH mass and stellar mass is significantly more consistent with the MBII prediction ($\sim0.3$~dex) favoring the hypothesis of AGN feedback being responsible for a causal link between the BH and its host galaxy.

In this study, we extend our previous work by adding recent measurements of hundreds of Sloan Digital Sky Survey (SDSS) quasars at $0.2<z<0.8$ based on wide and deep Hyper Suprime-Cam (HSC) imaging from the Strategic Subaru Program, and comparing the observational measurements with that from simulations. Furthermore, we extend the simulated quasar populations by including MBII, SAM, Illustris, TNG100, TNG300, and Horizon-AGN. This paper is structured as follows. In Sections~\ref{sec:observations} and~\ref{sec:simulations}, we describe our observed and simulated samples. A direct comparison is performed and the result is presented in Section~\ref{sec:result}. The concluding remarks are presented in Section~\ref{sec:con}.

\section{Observational data set}
\label{sec:observations}
The observed sample consists of 646 uniformly-selected SDSS quasars at $0.2<z<0.8$, imaged by Subaru/HSC \citep{Li2021a}, and 32 quasars at $1.2<z<1.7$ as imaged by~\hst~\citep[][hereafter D20]{Ding2020}. The latter are selected from three deep-survey fields, namely COSMOS~\citep{Civano2016}, (E)-CDFS-S~\citep{Lehmer2005, Xue2011}, and SXDS~\citep{Ueda2008}. Further details of these two samples and their measurements are given below. 

\subsection{SDSS/HSC sample}\label{sec:hsc}
A sample of $\sim$5000 type-1 SDSS quasars from the DR14 catalog~\citep{Paris2018} at $0.2<z<1$ has been imaged by the high-resolution Subaru Strategic Program (SSP) wide area survey~\citep{Aihara2019} using Hyper Suprime-Cam~\citep{Miyazaki2018}. With accurate PSF models in five optical bands {\it grizy}, two-dimensional quasar-host decompositions have been performed \citep[][]{Li2021a} to obtain the flux and color of each quasar's host galaxy. The state-of-the-art image modeling software {\tt lenstronomy} \citep{Birrer2015, Birrer2018, Birrer2021} is adopted to perform the modeling task. This approach is first developed by~\citet{Ding2020} and used to decompose the near-infrared emission of the HST sample (see next section). Having measured the host light in each band, the stellar mass of host galaxy is derived using spectral energy distribution (SED) fitting with CIGALE~\citep{Boquien2019}. Simulation tests are also performed to verify the fidelity of the \smass\ measurements. The statistical measurement error on \smass\ is at the $\sim$0.2~dex level. The values of \mbh\ are determined by~\citet{Rakshit2020} which are estimated based on the \hbeta-based measurements using the virial method~\citep{Peterson2004, Vestergaard2006}. The typical error of \mbh\ are estimated to be 0.4 dex. The mass ranges for the entire sample are log(\smass) $\in$ [9.0, 11.5] $M_{\odot}$, and log(\mbh) $\in$  [6.5, 10.0] $M_{\odot}$).
 We refer the reader to~\citet{Li2021a} in the Section~4.2 for more details.

To avoid any potential biases related to the selection of the quasars, \citet{Li2021b} isolated 877 sources which are uniformly selected based on their PSF-magnitudes, color cuts using single-epoch SDSS photometry and the value of the measured \smass. Specifically, we use the {\it ugri} color-selected sample (228 sources) from SDSS I/II~\citep{Richards2002}, and the CORE sample from SDSS BOSS (408 sources) and eBOSS (241 sources) surveys~\citep{Ross2013, Myers2015} (hereafter the uniform sample). These samples are initially selected based on PSF-magnitude cuts of $15 < i < 19.1$ (for {\it ugri}), and $i > 17.8$ and $g, r < 22.0$ (for CORE). Furthermore, a limit on \smass\ is set to assure the detection of the host, especially since the rate and accuracy of detection is higher when \smass\ is increasing, resulting in a final sample of 646 quasars. These selections will be adopted in an equivalent manner to the simulated samples to mitigate selection effects thus allowing the fair comparison.

\subsection{HST sample}

A sample of 32 HST-observed AGN systems across the redshift range $1.2<z<1.7$ are selected from three deep-survey fields (COSMOS, (E)-CDFS-S, and SXDS). The HST/WFC3 IR camera is used to obtain the high-resolution imaging data (HST program GO-15115, PI: John Silverman) with six position dither pattern and a total exposure time $\sim$2348~s. The filters F125W ($1.2<z<1.44$) and F140W ($1.44<z<1.7$) were employed, according to the redshift of each target to bracket the 4000~\AA~break.  The AGN images are analyzed and decomposed to infer the host galaxies fluxes using the approach developed by D20 based on {\tt lenstronomy}. The HST ACS/F814W imaging data for 21/32 of the AGNs is also used to infer the host color. The results show that stellar templates of 1 and 0.625~Gyr can match the sample color at $z<1.44$ and $z>1.44$, respectively (see Figure 5 in D20). These best-fit models are used to estimate the stellar masses of the host galaxies. \mbh\ is determined by \citet{Schulze2018} using near-infrared spectroscopic observations of the broad \halpha\ emission line with the recipe provided by~\citet{Vestergaard2006}, in a consistent manner to that adopted for HSC sample. The mass ranges for the HST sample are log(\smass) $\in$ [9.5, 11.0] $M_{\odot}$, and log(\mbh) $\in$  [7.5, 9.0] $M_{\odot}$).
 We refer the reader to D20 for a more detailed description of the analysis. 

The measurements of the \mbh-\smass\ relations for both the HST and HSC samples are obtained with a consistent approach. Thus, we expect the measurement errors of these two samples to be at a comparable level (i.e., $\Delta$\mbh$=0.4~$dex, $\Delta$\smass$=0.2~$dex). 
Indeed, the two samples are consistent with a lack of evolution in the mass ratio~\citep[see Figure 6 of][]{Li2021b}, even though the sample selection is slightly different.

\begin{deluxetable*}{lccccc}
\tablecaption{Key characteristics of hydrodynamic simulations used in this study.\label{tab:sim_sum}}
\tablewidth{0pt}
\tablehead{
\colhead{Simulation} & MBII & Illustris &  TNG100 & TNG300 & Horizon-AGN
}
\startdata
Box sizes $(\mathrm{cMpc})^3$  & $142.7^3$ & $(106.5)^3$ & $(111)^3$ & $(302)^3$ & $(142)^3$ \\
Particles  & $2\times1792^3$ & $2\times1820^3$ & $2\times1820^3$ & $2\times2500^3$ & $\sim2\times1024^3$ \\
\cline{1-6}
{\bf Mass Resolution} \\
 Dark matter &$1.57\times10^7$ & $6.26\times10^6$ & $7.5\times10^6$ & $5.9\times10^7$ & $8\times10^7$ \\
 Baryonic matter &$3.14\times10^6$ & $1.26\times10^6$ & $1.4\times10^6$ & $1.1\times10^7$ & $2\times10^6$ \\
\cline{1-6}
{\bf AGN Feedback} & \multicolumn{2}{c}{(feedback efficiency $\times$ radiative efficiency)} \\
High acc. mode & 0.05 $\times$ 0.1 & 0.05 $\times$ 0.2 & 0.1 $\times$ 0.2 & 0.1 $\times$ 0.2 & 0.15 $\times$ 0.1 \\
Low acc. mode  & -- & 0.35 $\times$ 0.2 & $\leq0.2 \times$ 0.2 & $\leq0.2 \times$ 0.2 & 1 $\times$ 0.1 \\
Transitions btw. modes & -- & 0.05 & \multicolumn{2}{c}{min[$2\times10^{-3}(\frac{M_{\rm BH}}{10^8M_{\odot}})^2$, 10\%]} & 0.01 \\
\cline{1-6}
AGN fueling mechanism   & ${4\pi G^2 M_{BH}^2 \rho}/{(c_s^2+v_{BH}^2)^{3/2}}$ & $\alpha \dot{M}_{Bondi}$ & $\dot{M}_{Bondi}$&$\dot{M}_{Bondi}$ & $\alpha\dot{M}_{Bondi}$ \\
Maximum accretion rate & 2 $\times$Edd. acc. rate & Edd. acc. rate & Edd. acc. rate & Edd. acc. rate & Edd. acc. rate\\
\enddata
\tablecomments{In the penultimate row, $\alpha$ is the boost factor. For Illustris, $\alpha = 100$; for Horizon-AGN, $\alpha =$ max$[(\rho/\rho_0)^2, 1]$. $\dot{M}_{Bondi}= 4\pi G^2M_{BH}^2 \rho/c_s^3$.
The `$\leq$' sign in the low accretion feedback mode of TNG means that the feedback efficiency follows a distribution with a maximum value of 0.2. Note that the radiative efficiencies are different in various simulations. Thus, different values are used for calculating the luminosity.
}
\end{deluxetable*}

\section{Simulations and comparison strategy}
\label{sec:simulations}
We introduce the simulation samples that are adopted in this study. All are based on larger-scale cosmological simulations, except the Semianalytic Model (SAM; see Section~\ref{subsec:SAM}). In Table~\ref{tab:sim_sum}, we summarized the key elements for each hydrodynamic simulation being considered. We note that each simulation adopts either a  \cite{2003PASP..115..763C} or a \cite{1955ApJ...121..161S} initial mass function (IMF). To make self-consistent comparisons, we ensure that the adopted IMFs for both the simulations and the observations are consistent. The assumed IMF is needed to obtain the mass-to-light ratio and convert the observed luminosity to stellar mass to perform the comparisons. For numerical simulation, the mass is the base material, which is IMF independent.

\subsection{{\tt MassiveBlackII} (MBII)}\label{subsec:MBII}
MBII is a high-resolution cosmological hydrodynamic simulation that has a box size of $(142.7~\mathrm{cMpc})^3$
and $2\times1792^3$~(i.e., dark matter + gas) particles. 
The simulation is based on smooth particle hydrodynamic (SPH) code \texttt{P-GADGET}, a hybrid version of the parallel code {\tt GADGET}~\citep{2005MNRAS.364.1105S}. The base cosmology parameters are based on the WMAP7 results~\citep{2011ApJS..192...18K}. For dark matter and gas, the mass resolutions are $1.57\times 10^7~M_{\odot}$ and $3.14\times 10^6~M_{\odot}$, respectively. The simulation includes a full modeling of gravity plus gas hydrodynamics, with a wide range of subgrid recipes to model the star formation~\citep{2003MNRAS.339..289S}, BH growth, and the feedback process~\citep{2005Natur.433..604D}. 

To model supermassive BH, the initial seed with mass $5\times 10^{5}~M_{\odot}/h$ is inserted into halos of mass $\gtrsim 5\times 10^{10}~M_{\odot}/h$. Once seeded, BH growth via gas accretion is assigned at a rate of $\dot{M}_{BH}={4\pi G^2 M_{BH}^2 \rho}/{(c_s^2+v_{BH}^2)^{3/2}}$ where $\rho$ and $c_s$ are the density and sound speed of the interstellar medium (ISM) gas at cold phase; $v_{BH}$ is the relative velocity between the BH and its surrounding gas. Note that unlike several previous works, the accretion rate in MBII adopt the prescription in~\citet{Pelupessy2007} which does not include any artificial boost factor. The accreted gas is released as radiation at a radiative efficiency of 10\%. A fraction of 5\% of the radiated energy thermally couples to the surrounding gas as BH (or AGN) feedback~\citep{2005Natur.433..604D}. A mildly super-Eddington (two times Eddington rate) is allowed. Due to resolution limitations, BH dynamics cannot be self-consistently modeled in the simulations. Two BHs are considered to be merged when their separation distance is below the simulation spatial resolution (i.e., the SPH smoothing length) and their relative speeds are lower than the local sound speed of the medium.

Halos are identified using a friends-of-friends (FOF) group finder~\citep{1985ApJ...292..371D}. Galaxies are identified with the stellar matter components of subhalos; these subhalos are identified using {\tt SUBFIND} within the halos~\citep{2005MNRAS.364.1105S}. As a common practice, the stellar mass is obtained by using a 3D spherical aperture of 30~kpc to represent the observed stellar mass.
We adopt this definition of stellar masses for all larger-scale cosmological simulations described in the following sections, except for Horizon-AGN which use total mass (see Section~\ref{subsec:Horizon}) for details.  Using this definition, \citet{Pillepich2018} has shown that the corresponding stellar mass function is consistent with the observational measure. Even more, the stellar mass using this 3D aperture can achieve good agreement to those measured within the Petrosian radii in observational studies~\citep{Schaye2015}. For further details of MBII simulation, we refer the reader to~\citet{Khandai2015}.

\subsection{Illustris}
The Illustris Project is another large scale hydrodynamics simulation, introduced in~\citet{Genel2014, 2014MNRAS.444.1518V, 2014Natur.509..177V, Sijacki2015, Nelson2015}. The simulation consist of a volume of (106.5 cMpc)$^3$~(slightly smaller than MassiveBlack II), 
and was run with the moving Voronoi mesh code {\tt Arepo}~\citep{2010MNRAS.401..791S} with a base cosmology adopted from WMAP9 results~\citep{2013ApJS..208...19H}. Besides gravity and gas hydrodynamics, the simulation calculates the astrophysical processes ~\citep{2013MNRAS.436.3031V, 2014MNRAS.438.1985T} that includes gas cooling and star formation~(with a density threshold of 0.13 cm$^{-3}$, \citealt{2003MNRAS.339..289S}), stellar evolution and chemical enrichment, kinetic stellar feedback by SNe activity, BH growth~(accretion and merging), and AGN feedback.

BHs are seeded with an initial mass of $1 \times 10^5~M_{\odot}/h$ when a halo exceeds a mass of $5 \times 10^{10}~M_{\odot}/h$. BHs then grow via accretion described by the Eddington limited Bondi-Hoyle-Lyttleton formalism ($\alpha4\pi G^2M_{BH}^2 \rho/c_s^3$), as well as mergers with other BHs. The boost factor $\alpha=100$ is introduced to account for the unresolved multiphase ISM~\citep{Springel2005, 2009MNRAS.398...53B}, which is otherwise expected to underestimate the density around the BHs. Lastly, accreting BHs radiate with a bolometric luminosity given by $\epsilon_r \dot{M}_{BH}c^2$, where $\dot{M}_{BH}$ is the mass accretion rate and $\epsilon_r=0.2$ is the radiative efficiency.

The AGN feedback consists of three components, namely quasar-mode, radio-mode and radiative feedback. In the quasar-mode which holds for BHs with Eddington ratio $>0.01$, the AGNs deposit $5\%$~(quasar-mode feedback efficiency) of their released energy into the surrounding gas as thermal energy. For Eddington ratios $<0.01$, the AGN feedback is in radio-mode where the thermal energy is released as hot bubbles with a radius of $\sim$ 100~kpc at the intervals between which the BH mass grows by a fixed fraction. 
The energy of the bubbles is given by $\epsilon_m \epsilon_r \delta M_{BH} c^2$ where $\delta M_{BH}$ is the change in BH mass within the last time interval, and $\epsilon_m=0.35$ is the radio-mode feedback efficiency. Lastly, the radiative feedback mode is implemented by modifying the heating and cooling rates of the gas in the presence of radiation from all surrounding AGN. Halos and galaxies are identified similar to those of MBII. As in MBII, a 3D 30~kpc spherical aperture is used to obtained the galaxy stellar mass. 

\subsection{IllustrisTNG}
{\it The Next Generation Illustris Simulations} (IllustrisTNG, hereafter, TNG)~\citep{2018MNRAS.475..676S, Pillepich2018, Naiman2018, Marinacci2018, Nelson2018} are a suite of magnetohydrodynamical simulations of galaxy formation in large cosmological volumes. It builds upon the scientific achievements of the Illustris simulation with improvements upon Illustris by 1) extending the mass range of the simulated galaxies and haloes, 2) adopting an improved numerical and astrophysical modeling, and 3) addressing the identified shortcomings of the previous-generation simulations. Note that TNG simulations employ a modified version of the Bondi formalism, with $c_s$ explicitly including a B term for these magneto-hydro simulations.

The TNG100 and TNG300 have a volume of (100~cMpc)$^3$ and (300~cMpc)$^3$, respectively. The adopted cosmological parameters are updates by the Planck result~\citep{2016A&A...594A..13P}.
The gas cooling and star formation prescriptions are broadly similar to the Illustris model. However significant updates have been made to the stellar feedback model~(more details in \citealt{2018MNRAS.473.4077P}).
BH seeds with initial mass of $8 \times 10^5 M_{\odot}/h$ are placed in Dark matter halos with a mass exceeding $5 \times 10^{10} M_{\odot}/h$. Notably, the seed mass is one order of magnitude higher than in the Illustris simulation. The BH accretion also follows the Bondi-Hoyle-Lyttleton formalism, but without any boost factor~(unlike Illustris). Accreting BHs release energy with a radiative efficiency of 0.2~(same as Illustris). The inclusion of the magnetic fields can affect the relationship between the BHs and their host galaxies properties; the \mbh-\smass\ mean relation is higher with magnetic fields~\citep{2018MNRAS.473.4077P}. 

The AGN feedback occurs in thermal, radio, and radiative modes. For high accretion rates, the feedback implementation is the same as in Illustris i.e., thermal energy is injected in the surroundings of the accreting BHs. However, at low accretion rates, the feedback implementation is substantially different from Illustris. Instead of releasing hot bubbles, this feedback mode in TNG is purely kinetic. In particular, there is a directional injection of momentum along a randomly chosen direction~\citep{2017MNRAS.465.3291W, 2018MNRAS.479.4056W}. 
The transition between the two feedback modes is also different from Illustris, and is set by the minimum value of 0.1 and $2 \times 10 ^{-3} \times (M_{BH} / 10^8~M_{\odot})$. Additionally, the radiative feedback implemented in Illustris~(summarized in the previous section) is also present in TNG. Lastly, halo and galaxy identification, as well as calculation of galaxy stellar mass, is done in a similar manner to that of Illustris and MBII.

\subsection{Horizon-AGN}\label{subsec:Horizon}
The simulation Horizon-AGN~\citep{2014MNRAS.444.1453D, 2016MNRAS.463.3948D} has a volume of 142 cMpc$^3$ and was generated using the adaptive mesh refinement code {\tt Ramses}~\citep{2002A&A...385..337T} with a $\Lambda$CDM model based on WMAP7~\citep{2011ApJS..192...18K} cosmological results. The dark matter particle mass is $8\times 10^7 M_{\odot}$. The stellar particle mass is $2\times 10^6 M_{\odot}$ and the MBH seed mass is $10^5 M_{\odot}$. Adaptive mesh refinement is permitted down to $\Delta x=1$~kpc, and, if the total mass in a cell becomes greater than 8 times the initial mass resolution, it is performed in a quasi-Lagrangian manner. Collisionless particles (dark matter and star particles) are evolved using a particle-mesh solver with a cloud-in-cell interpolation.
The simulation includes gas cooling down to $10^4\, \rm K$ \citep{sutherland&dopita93}, and stochastic star formation. Stellar feedback is modeled as mechanical energy injection from Type Ia SNe, Type II SNe and stellar winds, with the metal enrichment from these sources.

Differing from simulations presented above, Horizon-AGN does not use a fixed threshold in the dark matter halo mass to seed BHs.  BHs are seeded with a mass of $10^5 M_\odot$ in cells, with gas density above $n_0$ and stellar velocity dispersion larger than $100 \,\rm km\,s^{-1}$. An exclusion radius is imposed so that no BH seed is formed at less than 50 ckpc from an existing BH. After $z = 1.5$, new BHs are prevented from forming. At these subsequent times, all the progenitors of the \smass$>10^{10} M_{\odot}$ galaxies at $z = 0$ should be formed and seeded with BHs~\citep{2016MNRAS.460.2979V}.  BH accretion is computed using the Bondi-Hoyle-Lyttleton formalism with a boost factor $\alpha = (\rho/\rho_0)^2$ when the density $\rho$ is higher than the resolution-dependent threshold $\rho_0$. Otherwise, the boost factor is fixed as unity~\citep{2009MNRAS.398...53B}.

Horizon-AGN includes two modes of AGN feedback. In the quasar mode ($f_{\rm Edd}>0.01$), thermal energy is isotropically released within a sphere of radius a few resolution elements. The energy deposition rate is $\dot{E}_{\rm AGN} = 0.015 \dot{M}_{\rm BH} c^2$. In the radio mode, energy is injected into a bipolar  outflow  with  a  velocity  of  $10^4\,\rm km\,s^{-1}$, to  mimic the  formation  of  a  jet.  The  energy  rate  in  this  mode is $\dot{E}_{\rm AGN} = 0.1 \dot{M}_{\rm BH} c^2$.  The  technical  details  of  BH  formation,  growth  and AGN  feedback  modeling  of  Horizon-AGN  can be found in~\citet{2012MNRAS.420.2662D}.

We identify galaxies applying the AdaptaHOP structure finder \citep{Aubert+04,Tweed+09} to the star particle distribution.  Galaxies are identified using a local threshold of 178 times the average matter density, with the local density of individual particles calculated using the 20 nearest neighbors. Only galaxies with more than 50 particles are considered. With this approach for Horizon-AGN, the adopted galaxy mass corresponds to the total stellar mass of a galaxy, which is different than the other hydrodynamic simulations (i.e., within a 3D 30~kpc spherical aperture). This definition of stellar mass is commonly used in Horizon-AGN, whose stellar mass function is known~\citep{2017MNRAS.467.4739K} to be in good agreement with observations.

\begin{figure}
\centering
\includegraphics[height=0.37\textwidth]{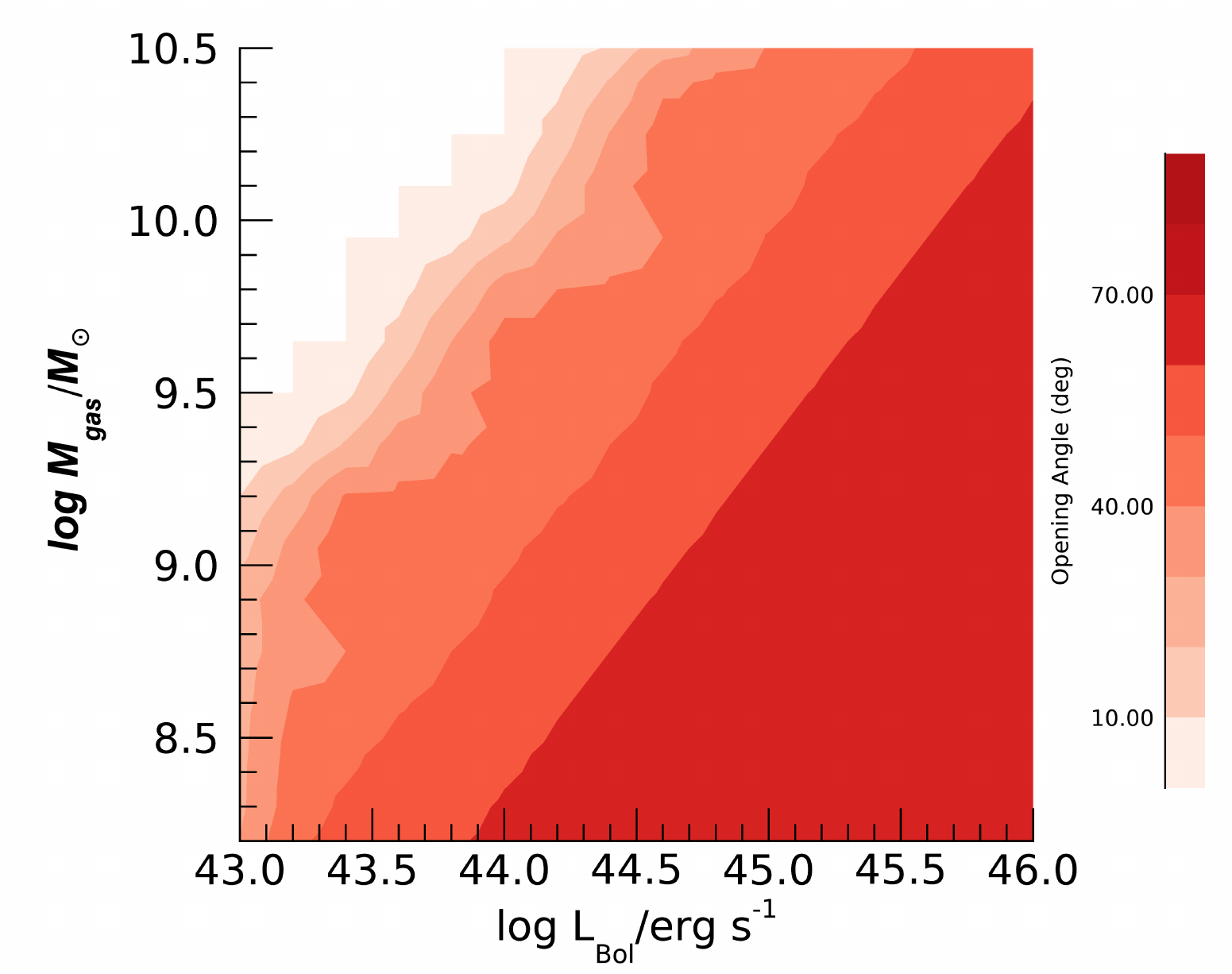}
\caption{\label{fig:SAM} 
Total gas content of galaxies as a function of AGN bolometric luminosity and jet opening angle in a new AGN feedback model incorporated into the SAM simulation.
}
\end{figure} 

\begin{figure*}
\centering
\begin{tabular}{c c}
{\includegraphics[trim = 0mm 0mm 65mm 10mm, clip, height=0.4\textwidth]{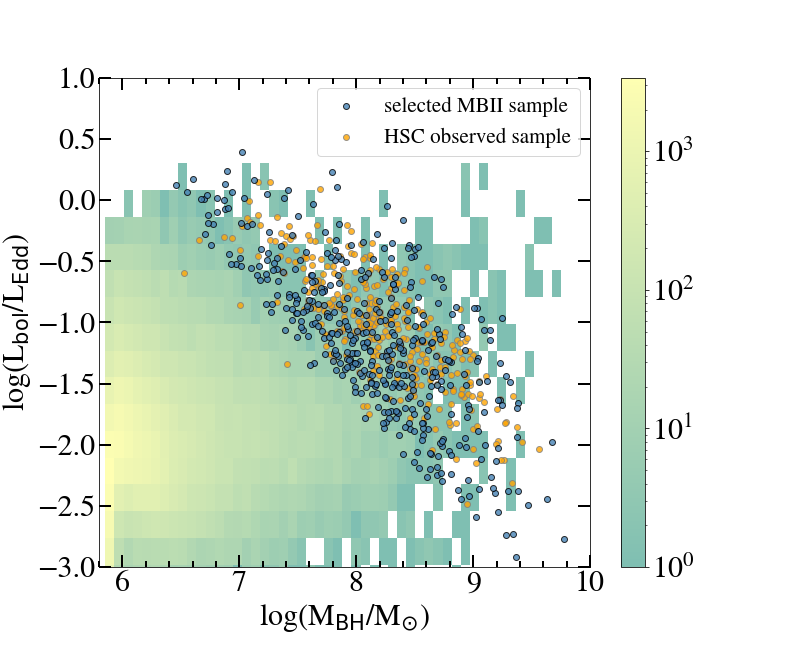}}
{\includegraphics[trim = 0mm 0mm 20mm 10mm, clip,height=0.4\textwidth]{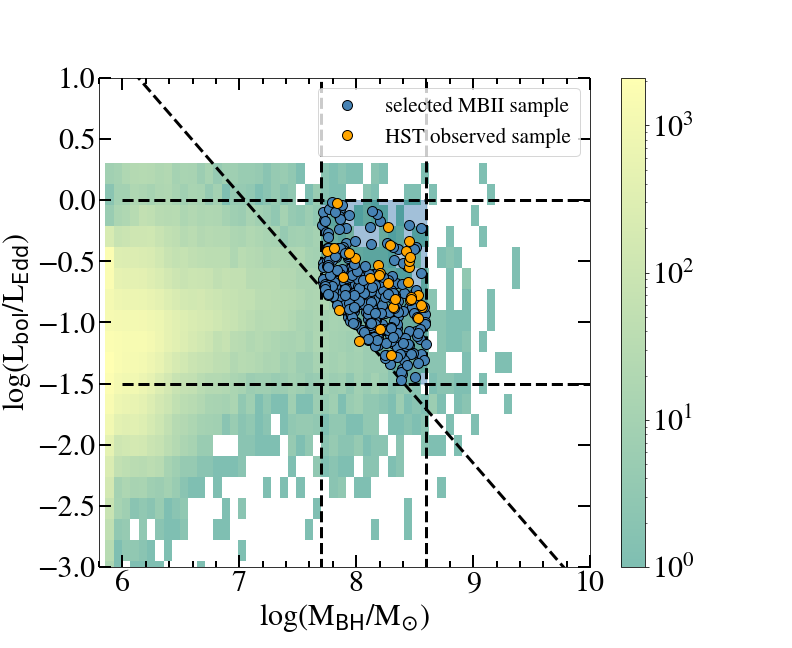}}
\end{tabular}
\caption{\label{fig:selection}Demonstration of AGN selection using MBII. {\it left}: distribution of \mbh\ and Eddington ratio for the full (colored squares) MBII sample and individual objects meeting the observed selection criteria (blue circles) for those at redshift $z=0.6$. A matched HSC sample is shown by the orange data points. The light-green background cloud shows the {\it intrinsic} simulated number density in this parameter space. Note that this is the first step of sample selection. We further use an AGN magnitude cut to assure that the simulation sample has a similar $L_{\rm bol}$ distribution (e.g., see Figure~\ref{fig:offsets_nochange}-left) and \mbh\ distribution with the observations. We then use the same \smass\ cut to construct the final sample.
{\it right}: similar to the left panel, but presenting the impact of selection on the HST sample.
}
\end{figure*}

\subsection{Semi-analytic Model (SAM)}\label{subsec:SAM}
We highlight the main points of the simulation with respect to our study; for more detail, a full description of the SAM can be found in~\citet{Menci2016} which is based on an earlier semi-analytic model introduced in~\citet{Menci2014}. The specific version adopted here differs from the one presented in the above papers since it implements a new, detailed description of AGN feedback, as discussed in detail below.

For dark matter halos that merge with a larger halo, the impact of dynamical friction is assessed to define whether the halo will survive as a satellite or sink to the center of the dominant galaxy which increases its mass. The binary interactions (fly-bys and mergers), among satellite sub-halos, are  also described by the model. In each halo, we compute the fraction of gas which cools because of the atomic processes and settles into a disk~\citep{Mo1998}. The stars are converted from the gas through three channels: (1) quiescent star formation with long timescales: $\sim1$~Gyr; (2) starbursts following galaxy interactions with timescales $\lesssim 100$~Myr;  (3) the loss of angular momentum triggered by the internal disk instabilities causing the gas inflows to the center, resulting in stimulated star formation (as well as BH accretion). The stellar feedback is also considered by calculating the energy released by the supernovae associated with the total star formation, which returns a fraction of the disk gas into a hot phase. 
A $\Lambda$CDM power spectrum of perturbations with a total matter density parameter $\Omega_0=0.3$, a baryon density parameter 
$\Omega_b=0.04$, a dark energy density parameter $\Omega_\Lambda=0.7$, and a Hubble constant $h$=0.7 is adopted.

{We assume BH seed $M_{seed}=100\,M_{\odot}$~\citep{Madau2001}  to be initially present in all galaxy progenitors at the initial redshift $z=15$. This constitutes an approximate way of rendering the effect of the collapse of Population III stars. However, the detailed value of $M_{seed}$ has a negligible impact on the final BH masses as long as they remain in the range $M_{seed}=50-500\,M_{\odot}$.
}

The BH accretion is assumed to follow from the gas instabilities resulting from either galaxy interactions or disk instabilities, and is thus related to star formation channels 2-3: a fraction of the cold gas destabilized during galaxy encounters and through disk instabilities is in fact accreted onto the central BHs (with the remaining fraction fueling star formation channels 2-3 described above). Such fractions and the corresponding timescales for accretion are computed as described in~\citet[][their Section 3]{Menci2014}.

The SAM adopted here implements a new and improved  model for the AGN feedback with respect to the previous versions~\citep{Menci2008}. In both versions, the basic assumption is that fast winds with velocity up to $10^{-1}c$ observed in the central regions of AGNs~\citep{Chartas2002, Pounds2003}  result in  supersonic outflows that compress the gas into a blast wave terminated by a leading shock front. This  moves outward with a lower but still supersonic speed, and sweeps out the surrounding medium. However, while in the earlier version of the SAM~\citep{Menci2016} the blast wave is assumed to expand into an isotropically distributed medium, in the new description of AGN feedback~\citep{Menci2019} the full two-dimensional structure of the gas disk and of the expanding blast wave is followed in detail. The main physical difference is that in the new model the large density of gas along the plane of the disk causes the blast wave expansion to stall in such a direction, while it expands with large velocities in the vertical direction. The resulting strong dependence of the total (integrated over directions) outflow rate on the AGN luminosity $L_{AGN}$ and on the gas content of the galaxy $M_{gas}$ is shown in Figure~\ref{fig:SAM}. Such a new AGN feedback model has been tested in detail against a state-of-the-art compilation of observed outflows in 19 galaxies with different measured gas and dynamical masses~\citep{Fiore2017}, allowing for a detailed, one-by-one comparison with the model predictions. This well tested AGN feedback model allowed us to derive, for each simulated galaxy in the SAM,  the outflow expansion and the mass outflow rates in different directions with respect to the plane of the disc.

\subsection{Application of observational measurement error and selection effects}\label{subsec:add_obs_eff}
To make direct comparisons with observations, we add measurement errors and apply the equivalent selection to the simulated samples. We first inject random noise to the simulated catalog to mimic the scatter caused by measurement error. As mentioned above, \smass\ and \mbh\ for HSC and HST samples are measured with a similar approach; thus, their uncertainty levels are expected to be equivalent. We assume the following measurement uncertainties that are added as random noise: $\Delta$\mbh$ = 0.4~$dex, $\Delta$\smass$ = 0.2~$dex, and $\Delta L_{\rm bol} = 0.03~$dex. Since the selection of the simulation sample based on $L_{\rm bol}$ has limited effects in this study (see the discussion in Section~\ref{sec:dis} and Figure~\ref{fig:offsets_nochange} for details), the AGN variability correction is not considered.

We then apply restrictions on the noise-injected simulation to mimic selection effects as present in the observational data. Since the HSC and HST samples have their own selection function, we apply different selection criteria to the simulation as follows.

Comparing with the HSC sample: (1) The observed sample consists of type-1 AGN, and thus the simulated sample should match the relationship between \mbh-$L_{\rm bol}$ as seen in the HSC sample. We use MBII to demonstrate the importance of matching the sample selection (Figure~\ref{fig:selection}--{\it left}). (2) The $i$-band magnitudes of the AGN are bright (see Section~\ref{sec:hsc}). 
The specific selection is carried out as follows: a value of AGN $i$-band magnitude is chosen to make the sample selection such that the $L_{\rm bol}$ distribution is similar to the observations (see Section~\ref{sec:dis} for details).
Since the simulations do not provide the observed AGN magnitude, we adopt a simulated rest-frame magnitude or L$_{\rm 5100}$ and assume the quasar continuum as a single power-law with an index of $\alpha_\nu=-0.44$~\citep{2001AJ....122..549V} to calculate the observed $i$-band magnitude.
 (3)~Following the HSC selection, we require the \smass\ value to be above a certain level (according to their redshift) to assure an accurate measurement. Finally, the HSC sample is split into three redshift bins for making comparisons, which are $0.2<z<0.4$, $0.4<z<0.6$, and $0.6<z<0.8$.
 
Comparing with the HST sample: Simulated AGN systems are selected only when they match the  \mbh-$L_{\rm bol}$ targeting window, which is the same as the observational selection (see Figure~\ref{fig:selection}--{\it right} using MBII as an example). Note that the selection of the HST sample has a hard cut on the \mbh\ values (i.e., log(\mbh) between [7.7, 8.6] $M_{\odot}$). The HST sample covers the higher redshift range $1.2<z<1.7$, which is considered as a single redshift bin to make the comparison with the simulations at $z=1.5$.

\section{Results} \label{sec:result}
For comparison, the local scaling relation provided by  D20 (e.g., \mbh$=0.98$\smass$-2.56$, Chabrier IMF\footnote{Since different simulations adopt either a Chabrier or a Salpeter IMF, we use the local relation and \smass\ of the observational data that are based on the same IMF; thus, a comparison between the observations and simulations are self-consistent.}) is adopted as the fiducial relation to assess relative offsets and differences in dispersion with redshift. This local (\mbh-\smass) relation is derived by fitting measurements for 55 local galaxies as given in~\citet{Bennert++2011} and \citet{H+R04}.
For each sample, we focus on the \mbh\ residuals\footnote{The value of the slope for the local sample is close to 1, and thus if taking the \smass\ to calculate the residual for each system, the offset value remains the same.} (i.e., the offset to the local relation along the $y$-axis) and calculate their mean and standard deviation to make comparisons with observations.

\begin{figure*}
\centering
\includegraphics[trim = 40mm 60mm 30mm 90mm, clip,height=1.1\textwidth]{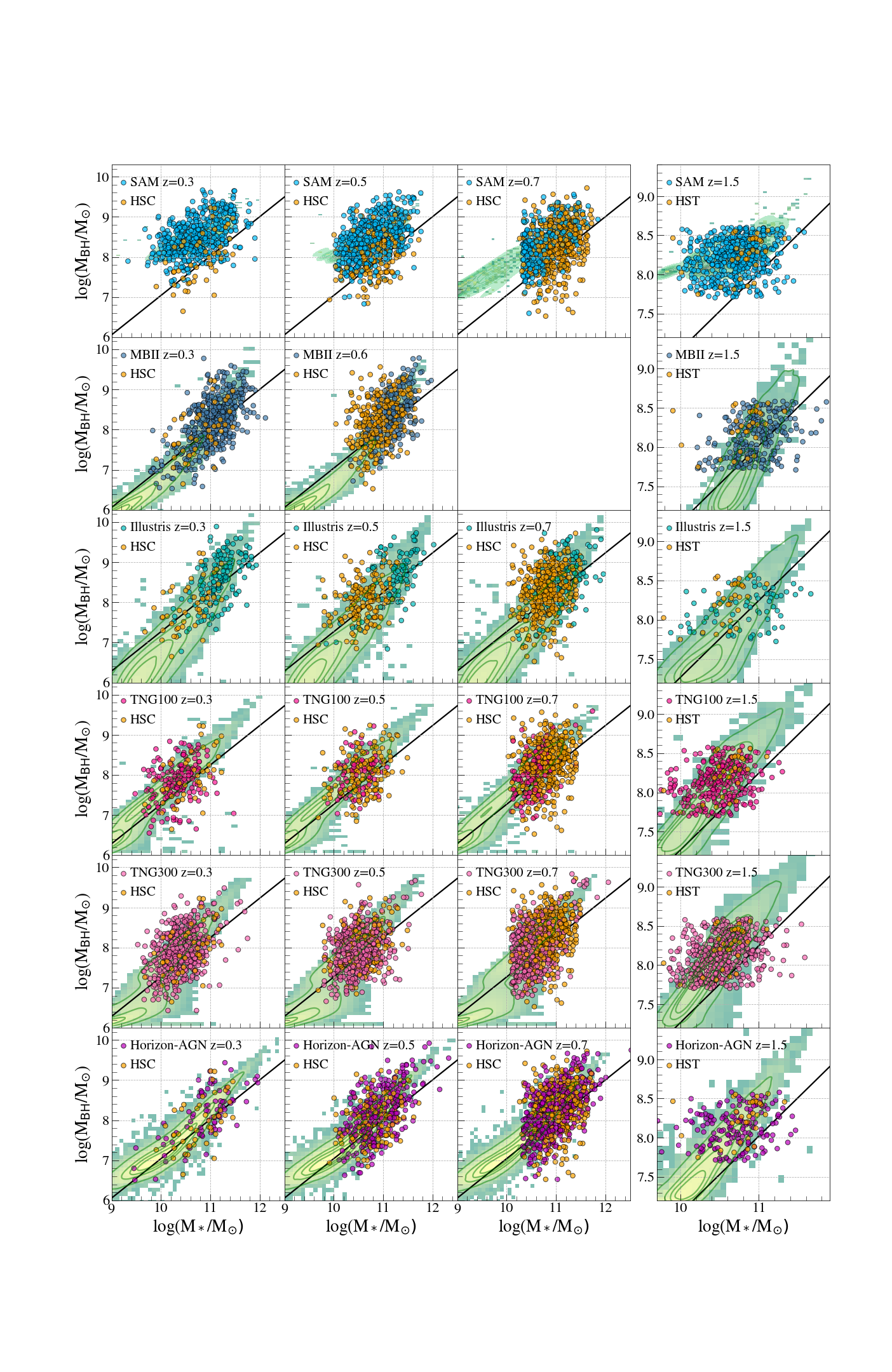}
\caption{\label{fig:comparsion} 
BH mass vs. stellar mass for both the observational (small orange circles) and simulated (small colored circles) samples. Each row pertains to a particular simulation as labeled. The panels, from left to right, show different redshift bins. The black line in each panel indicates the local relation adopted by~\citet{Ding2020}. The background cloud (in green and yellow with contours) shows the intrinsic simulation number density before injecting random noise and applying selection effects. The TNG100 and TNG300 appear to present similar results (see discussion section). We note that different samples adopt either a Chabrier or a Salpeter IMF for calculating the stellar mass; thus, the \smass\ values for the observations and the local relation are shifted appropriately.
}
\end{figure*}

In Figure~\ref{fig:comparsion}, we present the mass scaling relation \mbh--\smass\ for both the observations and simulations for direct comparison. For the simulated data, both the initial sample and that with observational effects applied (i.e., noise correction and selection) are shown. Note that offsets can occur for the simulated samples in three cases: an inherent offset, an offset due to selection, and offsets from added noise and selection. For the first, it has been recognized~\citep[e.g.,][Figure 2]{Habouzit2021}, and seen in our Figure~\ref{fig:comparsion}, that the  offset values vary over a range of stellar mass, and thus the mean and standard derivation do not represent the entire mass distribution.
In this work, we focus on the last two offset distributions, i.e., considering selection effects with and without noise correction. To aid in visualization of the differences among the various simulations, compared to the observed sample, we show the distribution of offsets (in terms of the $\Delta{\rm log}$\mbh) as histograms in Figure~\ref{fig:offsets}.  Each panel presents a different redshift range. In addition, the values of the central offset and scatter are presented 
in Table~\ref{tab:sum}, both before and after consideration of the noise. Comparisons with the observations are presented in the remainder of this section.

\subsection{Dispersion}

\begin{figure*}
\centering
\begin{tabular}{c c c c}
\hspace*{-0.4cm} 
{\includegraphics[height=0.4\textwidth]{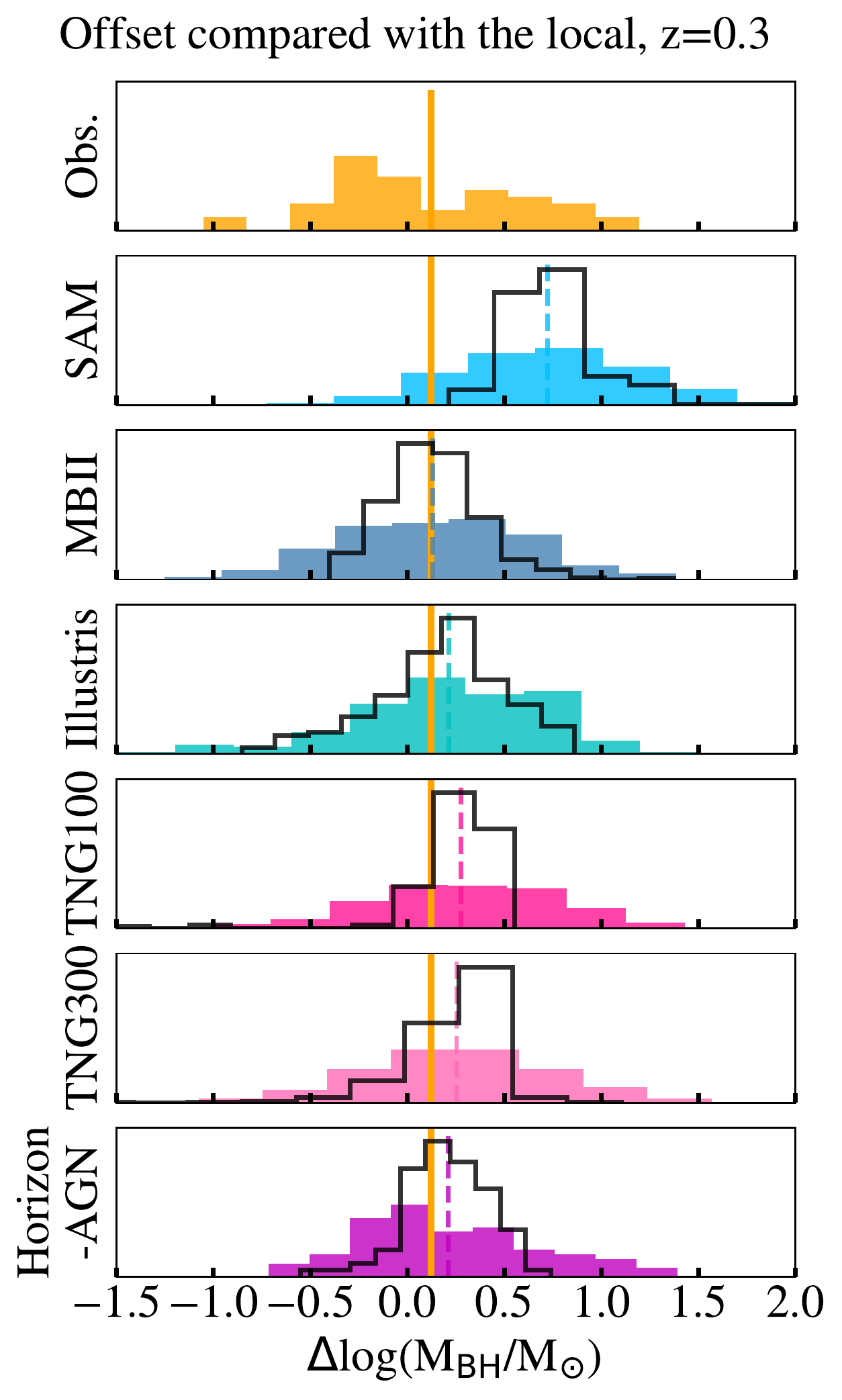}}&
\hspace*{-0.4cm} 
{\includegraphics[height=0.4\textwidth]{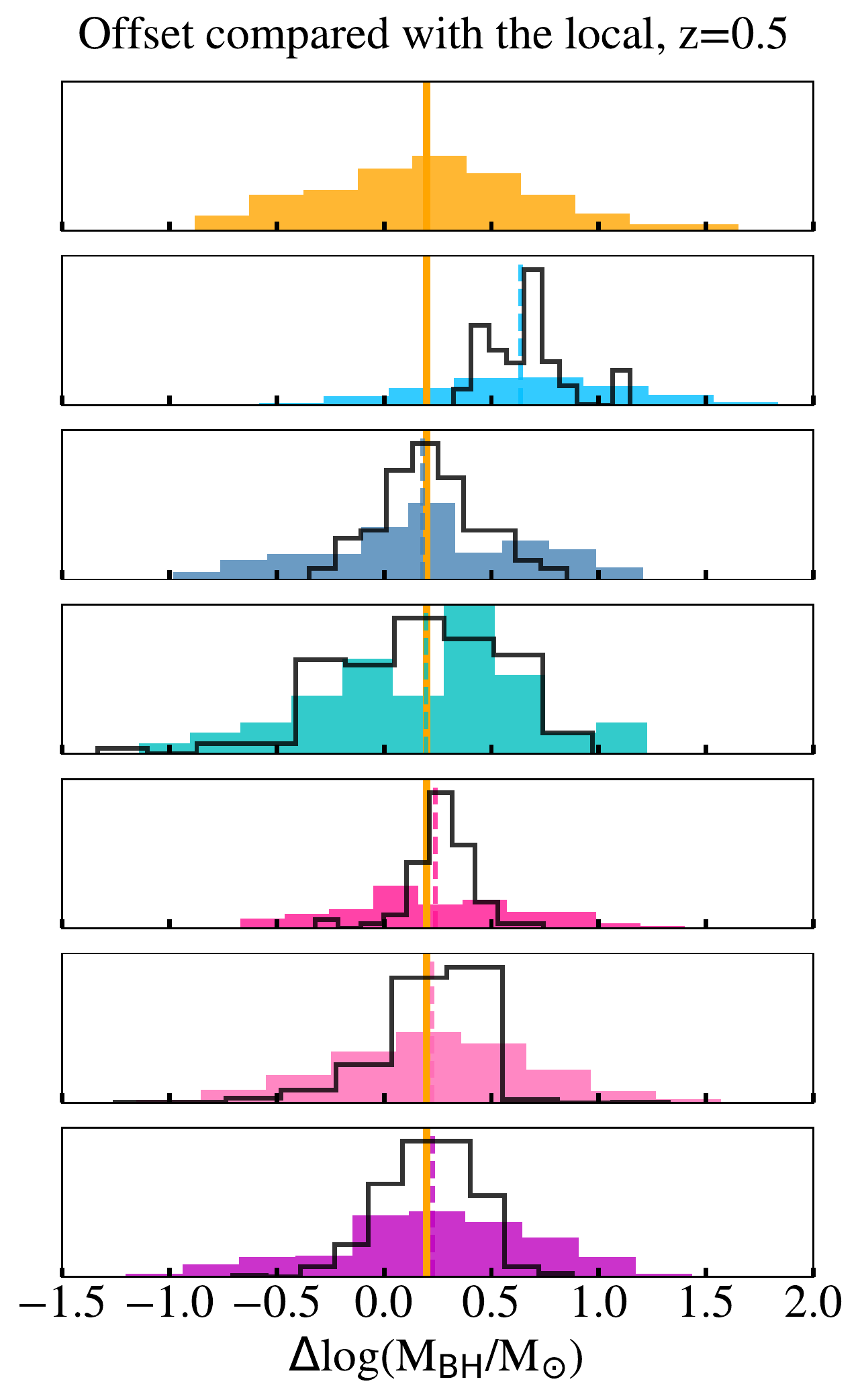}}&
\hspace*{-0.4cm} 
{\includegraphics[height=0.4\textwidth]{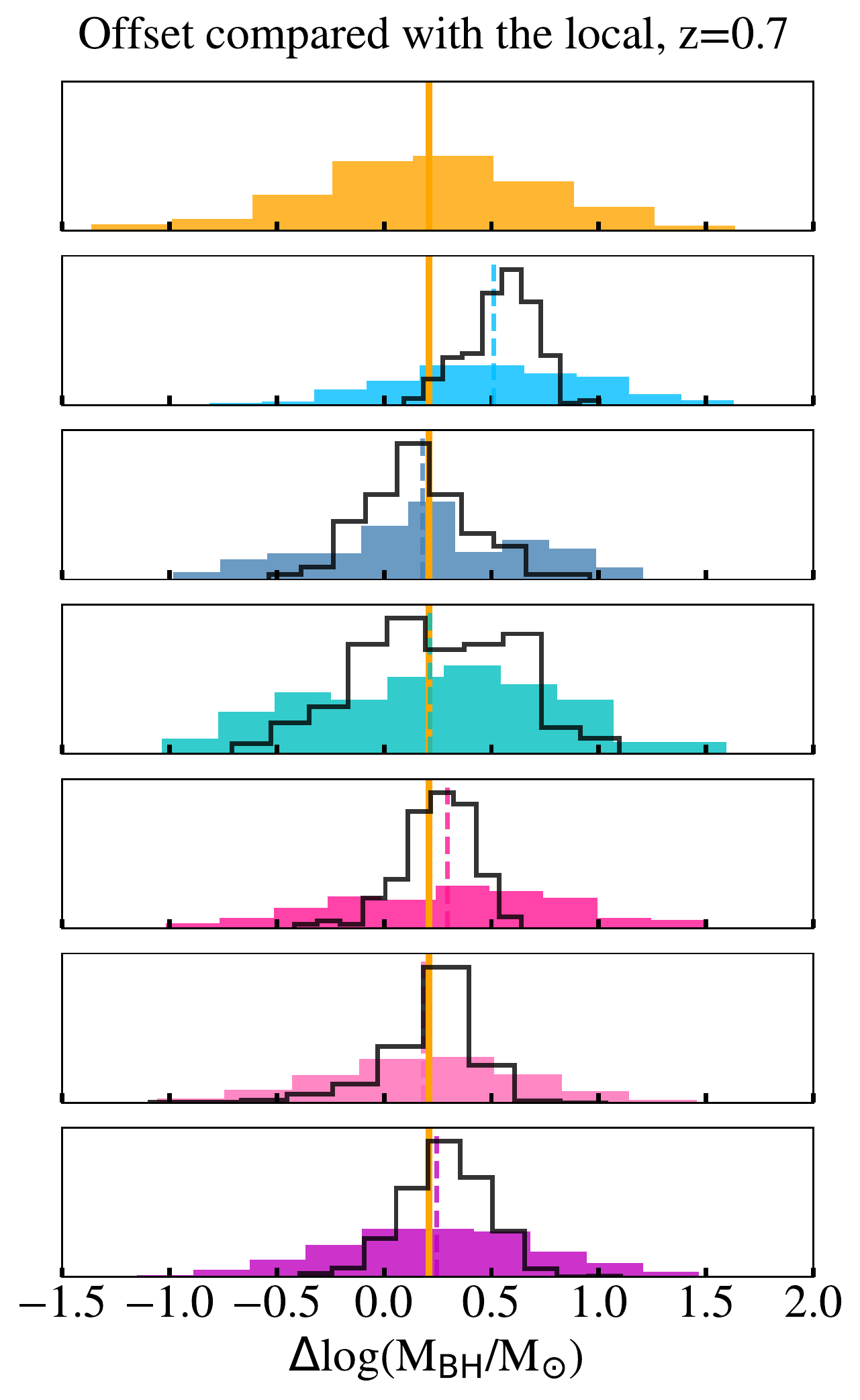}}&
\hspace*{-0.4cm} 
{\includegraphics[height=0.4\textwidth]{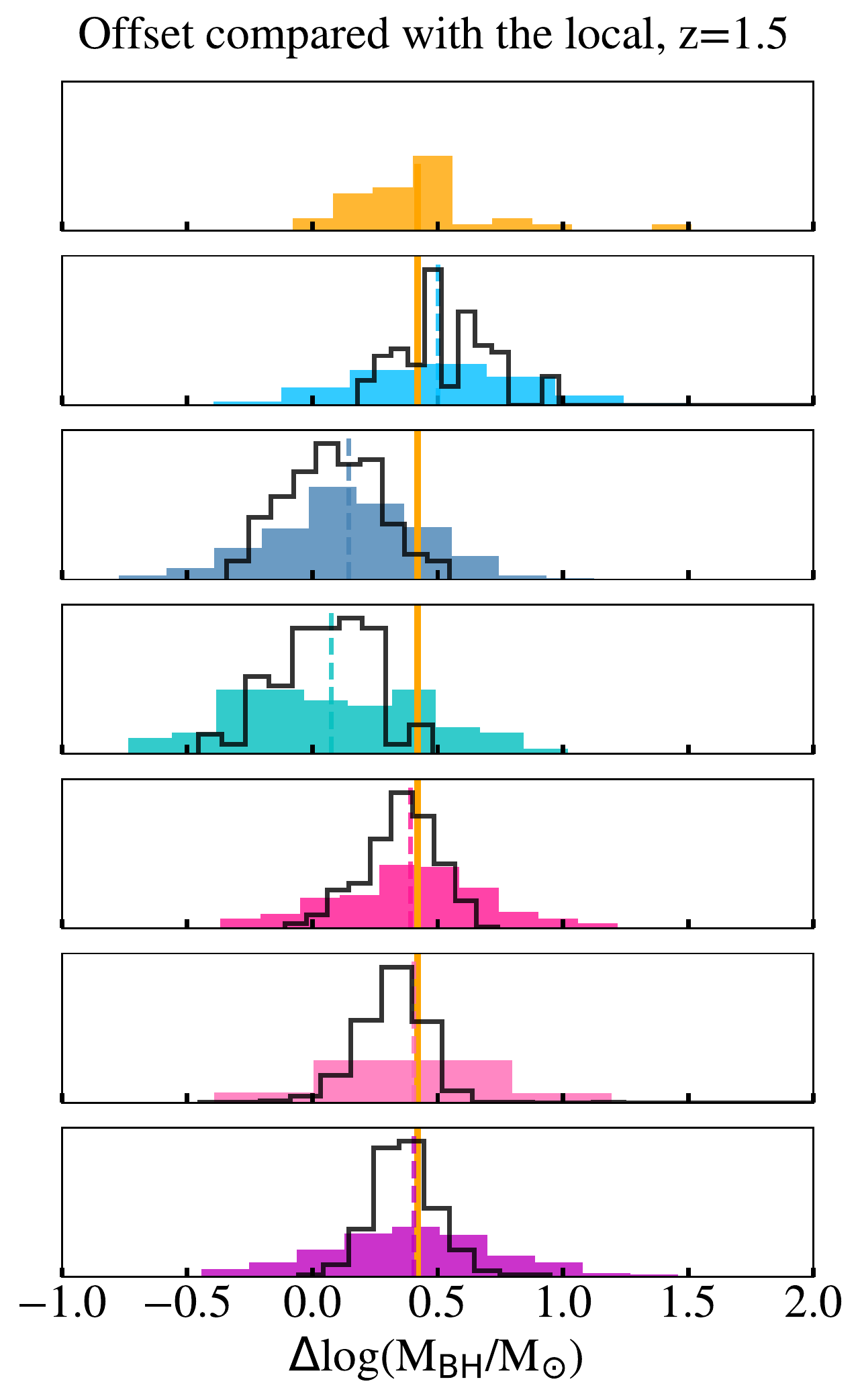}}\\
\end{tabular}
\caption{\label{fig:offsets} 
Colored histograms show the offset distributions for all simulated samples. The mean value and standard deviation of each are summarized in Table~\ref{tab:sum}. The vertical dashed lines indicate the corresponding mean value for each distribution. The mean values for the observed sample (i.e., yellow lines) are also given in each panel. To address the effect of noise, the offset distributions of the simulation without adding noise are also shown by the open black histograms.
For the MBII simulation, the sample at redshift 0.6 is used to compare with other samples at $z=0.5$ and $z=0.7$. 
}
\end{figure*}

Our results show that almost all simulations can produce scatter which is consistent with the observations across all redshifts examined (Figures~\ref{fig:comparsion}~and~\ref{fig:offsets}) --- for the simulated samples at $z<1$, this level of scatter is $\sim0.5$~dex, while at $z>1$, it is $\sim0.3$~dex. Note that the HST sample $z>1$ has a narrow selection window based on \mbh\ (see Figure~\ref{fig:selection}~bottom), causing the observed scatter to be smaller than that of the HSC sample at $z<1$. At all redshifts, we recognize that the observed scatter is dominated by measurement uncertainties in the data. 

\begin{figure}
\centering
\includegraphics[height=0.4\textwidth]{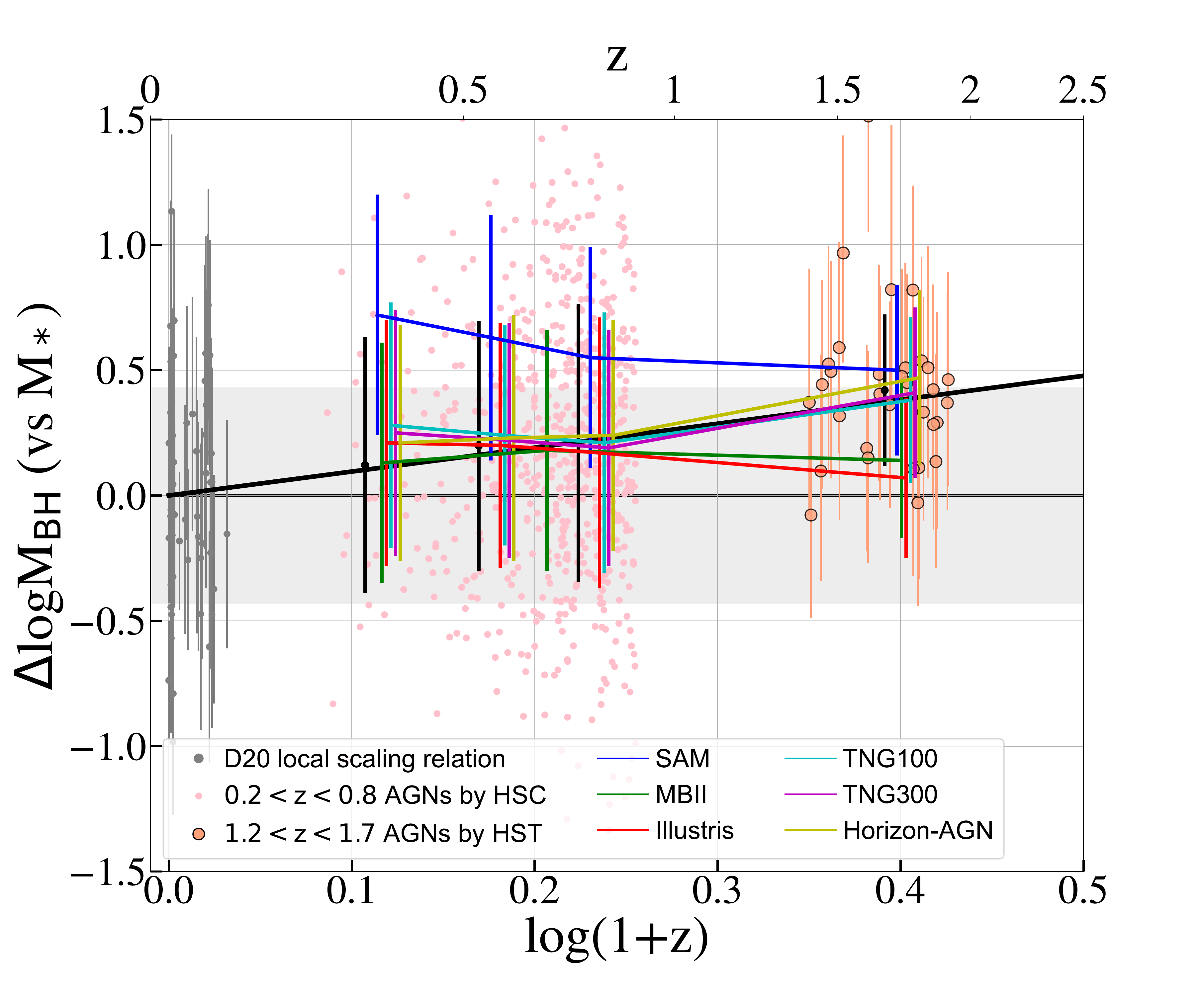}
\caption{\label{fig:offsets_vz} 
The {\it observed} evolution of $\Delta{\rm log}$\mbh\ as a function of redshift using both observational and simulated data. The black line shows the evolution by fitting the offset as a function of redshift. The black dots (with error bars) represent the mean (and standard deviation) values for the observations in four redshift bins. The \smass\ and \mbh\ ranges are different for each redshift bin, which are indicated in Figure~\ref{fig:comparsion}. The predictions from the numerical simulations, given in Table~\ref{tab:sum}, are presented by different colored symbols. The grey horizontal band illustrates the level of dispersion for the local sample (D20).
}
\end{figure}

An understanding of how much of the scatter derives from random noise can help us to determine the {\it intrinsic} scatter in the scaling relation. To this end, we measure the scatter of the simulation sample without injecting random noise but adopting the same selection window for both $z<1$ and $z>1$ samples to infer the central offset and scatter. We find that the intrinsic scatter is at a level of $\sim0.15-0.2$ dex for both $z<1$ and $z>1$ (see Table~\ref{tab:sum}). These levels are consistent with the {\it intrinsic} scatter as estimated using observation data alone~\citep{Ding2020, Li2021b}. Furthermore, the intrinsic scatter appears to be independent of redshift since the observations and simulations all follow the observed trend with redshift expected to be due to selection effects (Figure~\ref{fig:offsets_vz}). This suggests that the tight scaling relation may not be the result of a pure stochastic process, i.e., random mergers. However, the scatter is affected by sample selection, and thus these levels can only be taken as an approximation of the true intrinsic scatter.

\subsection{Global offsets}\label{subsec:offset}
We examine the offsets to understand whether the simulations deviate or not from the {\it observed} scaling relation, with particular attention to changes with redshift. Considering the values given in Table~\ref{tab:sum} and shown in Figure~\ref{fig:offsets_vz}, over the lower redshift range $z<0.6$, Illustris and Horizon-AGN predict {\it observed} \mbh\ offsets consistent with the observation data (at a level of $\lesssim0.1$~dex). At higher redshift $0.6<z<1.5$, the simulations SAM, TNG100, TNG300 and Horizon-AGN follow the {\it observed} evolution. These results are consistent with the Kolmogorov-Smirnov (KS) test performed using the offset distributions between each simulated sample and the observed sample --- the $p$-values are given in Table~\ref{tab:pvalue} showing that Horizon-AGN and Illustris have a good statistical match to the scaling relation at $z<0.6$ (i.e., $p$-value $> 0.1$), while the TNG100, TNG300 and Horizon-AGN simulation do well at $z>0.6$. 
Except for MBII and Horizon-AGN, we also see that the other simulations have mass offsets that are decreasing from $z=0.2$ to $z=0.6$, while the observation offsets increase with redshift up to $z\sim1.5$. However, this inconsistency is well below the $1\sigma$ scatter level.
Overall, we find that the mass correlation between supermassive BHs and their host galaxies is generally consistent between observations and simulations, with some subtle differences that are not at the level of concern for this present study.

\subsection{Trends with stellar mass}
\begin{figure*}
\centering
\begin{tabular}{c c}
\hspace*{-0.5cm} 
{\includegraphics[trim = 0mm 0mm 0mm 0mm, clip,
height=0.4\textwidth]{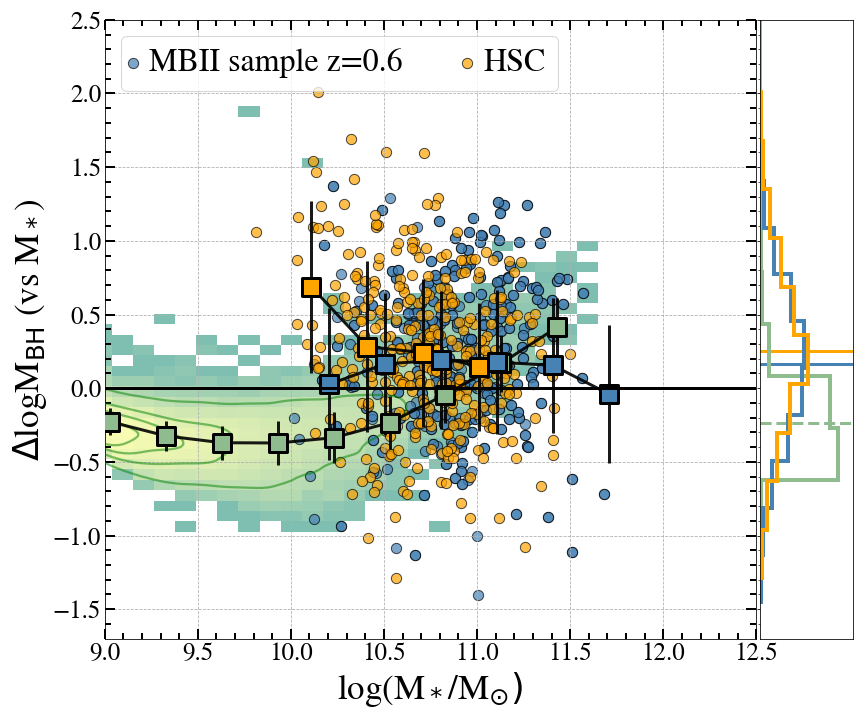}}&
\hspace*{-0.3cm} 
{\includegraphics[trim = 36mm 0mm 0mm 0mm, clip,
height=0.4\textwidth]{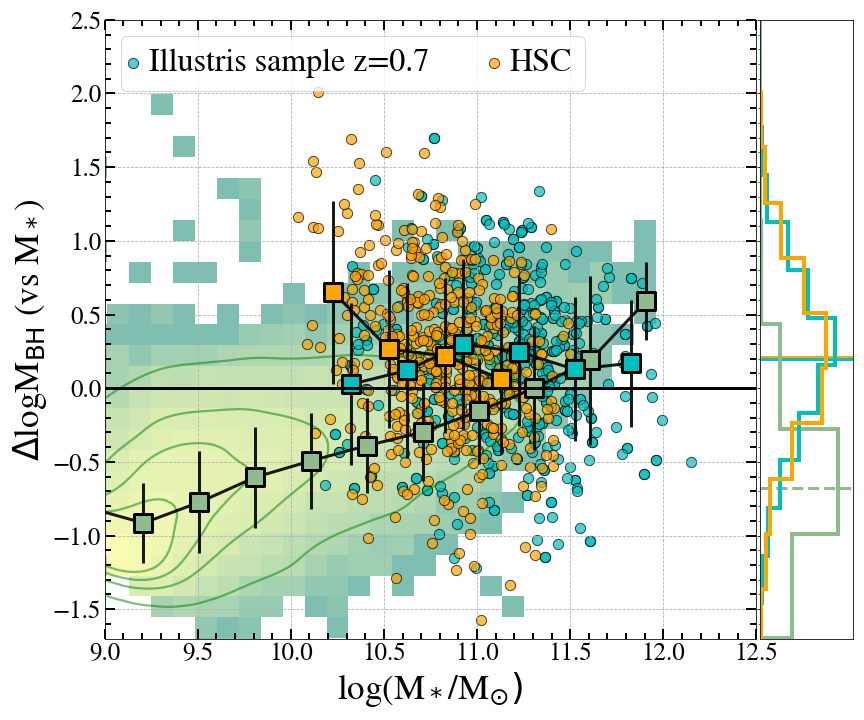}}\\
\hspace*{-0.5cm} 
{\includegraphics[trim = 0mm 0mm 0mm 0mm, clip,
height=0.4\textwidth]{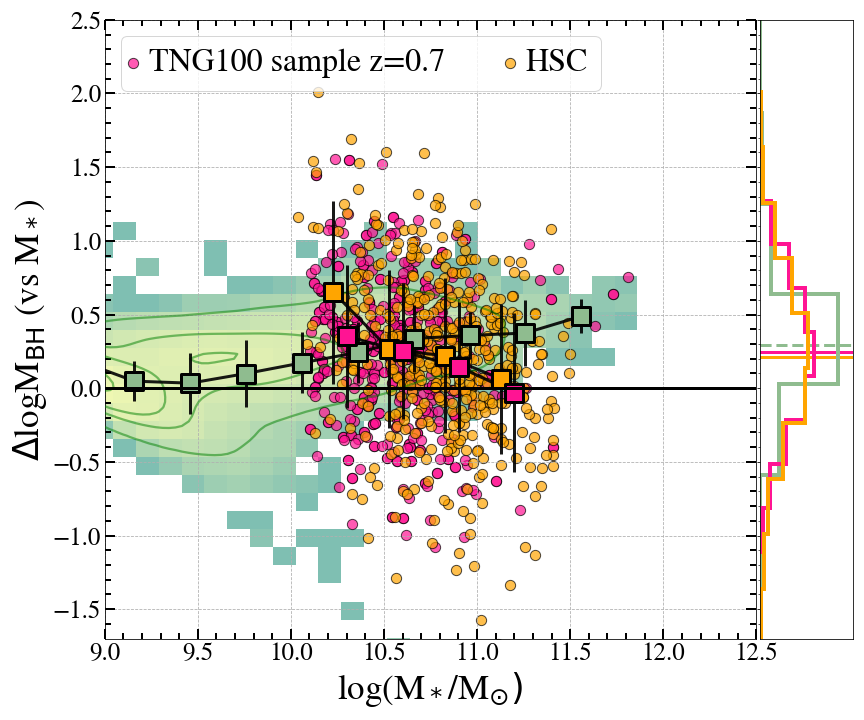}}&
\hspace*{-0.3cm} 
{\includegraphics[trim = 36mm 0mm 0mm 0mm, clip,
height=0.4\textwidth]{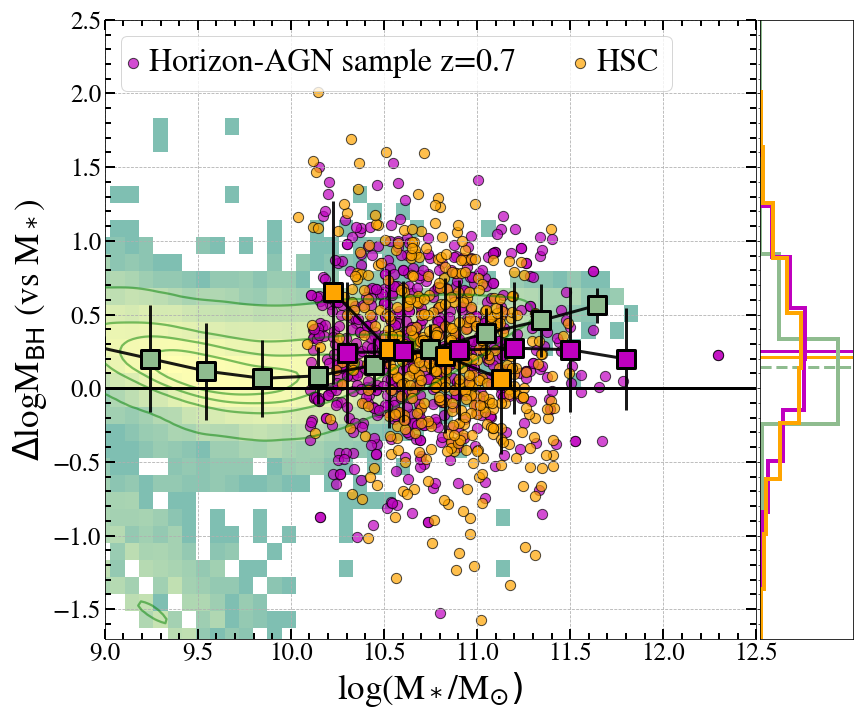}}\\
\end{tabular}
\caption{\label{fig:deltaMM} Comparisons of the offset of \mbh\ (to the D20 local scaling relation) as a function of stellar mass from observational data and the simulations at $z\sim0.7$. In each stellar mass bin (minimum number of objects is larger than 6), we give the mean and standard deviation of the offset values. To consider random noise, we use the average of ten realizations to calculate the mean and standard deviation in each bin. The histograms on the right show the offset distribution with lines marking the mean offsets for the observations and simulations using the full galaxy sample. The green color distributions with contours show the intrinsic simulated sample distribution without random noise and selection effect.
}
\end{figure*}

In Figure~\ref{fig:deltaMM}, we investigate how the offset values are correlated with stellar mass. Here we focus on the sample at $z\sim0.7$. The other redshift bins at $z<1$, where there is a large observation sample from HSC, show similar trends. We include the intrinsic values from the simulations in the figures to address how the observational effects (i.e., random noise and selection) change the observed scaling relations and offsets. First, considering the observed quasar sample (same in each panel), there is a trend for which BHs have masses further offset from their stellar mass with decreasing stellar mass. This trend is not seen in any of the simulations after noise, and selection effects have been applied. Given the level of uncertainties in the mean offsets of the observed sample, we do not try to interpret this trend any further in this study.

Interestingly, we notice that MBII and Illustris have BHs intrinsically undermassive relative to their galaxies at the lower masses that reach the D20 local scaling relation at higher masses. In contrast, TNG and Horizon-AGN have BHs slightly elevated from D20 local relation at most masses. These differences between simulations present two different scenarios, either one where the BHs come later, or coevolution with the two growing in tandem. Considering the former scenario, Illustris shows the strongest trend with stellar mass. In fact, after noise and selection are applied, the simulated sample exhibits very small offsets that agree remarkably well with the observed data. If the BHs are accurately characterized in the simulation, one interpretation is that the observations, including our HSC AGN sample, do not inform us of the true \mbh\ offsets as a function of \smass.  This result underscores the importance of taking into account errors and selection --- without accounting for those, one could erroneously interpret an apparent trend as evolution in the opposite sense as the true one. The most direct way to circumvent these issues is to probe lower masses (\smass\  $<10^{10}M_{\odot}$) using a more sensitive instrument, such as the James Webb Space Telescope~\citep{Habouzit2022}, across this redshift range \citep[see also][]{2011MNRAS.417.2085V}.

\begin{deluxetable*}{lccccc}
\tablecaption{Summary of the central offsets and scatters\label{tab:sum}}
\tabletypesize{\footnotesize}
\tablehead{
\colhead{Sample} &  \multicolumn{3}{c}{HSC Comparison} & \colhead{HST Comparison} \\
\cline{2-4} 
\colhead{} &  \colhead{$0.2<z<0.4$} & \colhead{$0.4<z<0.6$} & \colhead{$0.6<z<0.8$} & \colhead{$1.2<z<1.7$} & IMF
}
\startdata
Observation & (0.12$\pm$0.51) & (0.20$\pm$0.50)  & (0.21$\pm$0.56)  & (0.43$\pm$0.31) & \\
SAM & (0.75$\pm$0.22)$\rightarrow$(0.72$\pm$0.48) & (0.64$\pm$0.17)$\rightarrow$(0.63$\pm$0.49)  &(0.55$\pm$0.16)$\rightarrow$(0.55$\pm$0.44)  &  (0.52$\pm$0.18)$\rightarrow$(0.50$\pm$0.34) & Salpeter \\
MBII & (0.13$\pm$0.25)$\rightarrow$(0.13$\pm$0.48) & \multicolumn{2}{c}{[$z=0.6$] (0.22$\pm$0.23)$\rightarrow$(0.18$\pm$0.48)}  & (0.08$\pm$0.19)$\rightarrow$(0.14$\pm$0.31) & Salpeter\\
Illustris & (0.17$\pm$0.31)$\rightarrow$(0.21$\pm$0.49) & (0.17$\pm$0.38)$\rightarrow$(0.20$\pm$0.49)  &(0.32$\pm$0.36)$\rightarrow$(0.17$\pm$0.54)  &  (0.04$\pm$0.19)$\rightarrow$(0.07$\pm$0.32) &  Chabrier \\
TNG100 & (0.26$\pm$0.24)$\rightarrow$(0.28$\pm$0.49) & (0.27$\pm$0.15)$\rightarrow$(0.24$\pm$0.44)  &(0.25$\pm$0.17)$\rightarrow$(0.21$\pm$0.52)  &  (0.36$\pm$0.15)$\rightarrow$(0.38$\pm$0.33) & Chabrier \\
TNG300 & (0.25$\pm$0.22)$\rightarrow$(0.25$\pm$0.49) & (0.21$\pm$0.23)$\rightarrow$(0.22$\pm$0.47)  &(0.22$\pm$0.22)$\rightarrow$(0.19$\pm$0.47)  &  (0.32$\pm$0.16)$\rightarrow$(0.41$\pm$0.34) & Chabrier \\
Horizon-AGN & (0.21$\pm$0.22)$\rightarrow$(0.21$\pm$0.47) & (0.21$\pm$0.21)$\rightarrow$(0.23$\pm$0.49)  &(0.29$\pm$0.19)$\rightarrow$(0.24$\pm$0.46)  &  (0.37$\pm$0.13)$\rightarrow$(0.47$\pm$0.35) & Salpeter\\
\enddata
\tabletypesize{\normalsize}
\tablecomments{This table collects the comparison results of the \smass-\mbh\ correlations between different simulation at different redshift. The value shows the central position offset to the D20 local scaling relation and the scatters measured around the local relation after applying the offset. A positive offset means the \mbh\ value predicted by the simulation is higher than the local relationship measurement at fixed \smass\ value. The last column shows the corresponding IMF that was adopted to the local anchor to make a fair comparison with the observation. Note that for the observational data, the relative differences between local and high-$z$ measurements are not affected by the IMF assumptions.
For the MBII sample, the simulation does not produce the sample at $z=0.5$ or $z=0.7$, but rather at $z=0.6$. We use a Monte Carlo approach with 500 realizations to infer the uncertainties of the values in the table, finding that the uncertainties are within $\pm 0.03$.
}
\end{deluxetable*}

\section{Discussion} \label{sec:dis}
In this study, there are a few issues that may bias the results and thus need to be considered. First, the mass offsets are compared to the observed relation derived in the local universe (D20). In fact, the different simulations could have different mean relations at $z=0$~\citep[e.g.,][]{Habouzit2021}. As a result, the interpretation of the BH mass offsets of the simulations with redshift, anchored to the D20 local scaling relation, can be different if using the local relation of each individual simulation. In addition, the stellar mass of the Horizon-AGN sample is the total mass, while that for the other hydrodynamic simulations is determined within a 3D 30 kpc spherical aperture\footnote{We checked the use of total stellar mass for Illustris, TNG100, and TNG300 in these comparisons; the results are consistent with those based on a 30 kpc region.}. Therefore, in Figure~\ref{fig:offsets_vz_shiftz03}, we manually recalibrate the offsets of all simulations so that their mean value at $z=0.3$ is fixed to match that of the observations. This enables the evolutionary trend in the offsets of each sample to be clearer. In general, the updated results show that the evolution of the simulations is consistent with our interpretations in Section~\ref{subsec:offset}. This result can be expected since most of the simulations have demonstrated a good match to the observations at redshift $z=0.3$.

Regarding the survey volumes, we collected all available samples in the simulations to perform these comparisons. The final sample sizes, used to compare with observation, are not set by design. Despite that the observed samples are not volume limited and the simulated samples are not volume matched, fortunately, the volumes of all the simulations appear to be sufficient, i.e., all simulations have a similar number (or more) of data points to compare with the observations (see Figure~\ref{fig:comparsion}). In addition, similar results are found with the smaller- and larger-volumes simulations, e.g., TNG100 and TNG300, which reflect the fact that the simulation volume reaches a sufficient size to be effective for the comparisons in the observational plane. Even though, the intrinsic scatter of TNG100 and TNG300 (before the noise/selection) can differ.\footnote{We note that 
the subgrid models of the TNG simulation can change the intrinsic distributions at low redshift, especially for log~\smass$<10.5~M_{\odot}$ (see green contours in Figure 3 for the $z<0.7$ sample). }

\begin{figure*}
\centering
\begin{tabular}{c c c}
\hspace*{-0.4cm} 
{\includegraphics[trim = 25mm -10mm 20mm 15mm, clip, height=0.3\textwidth]{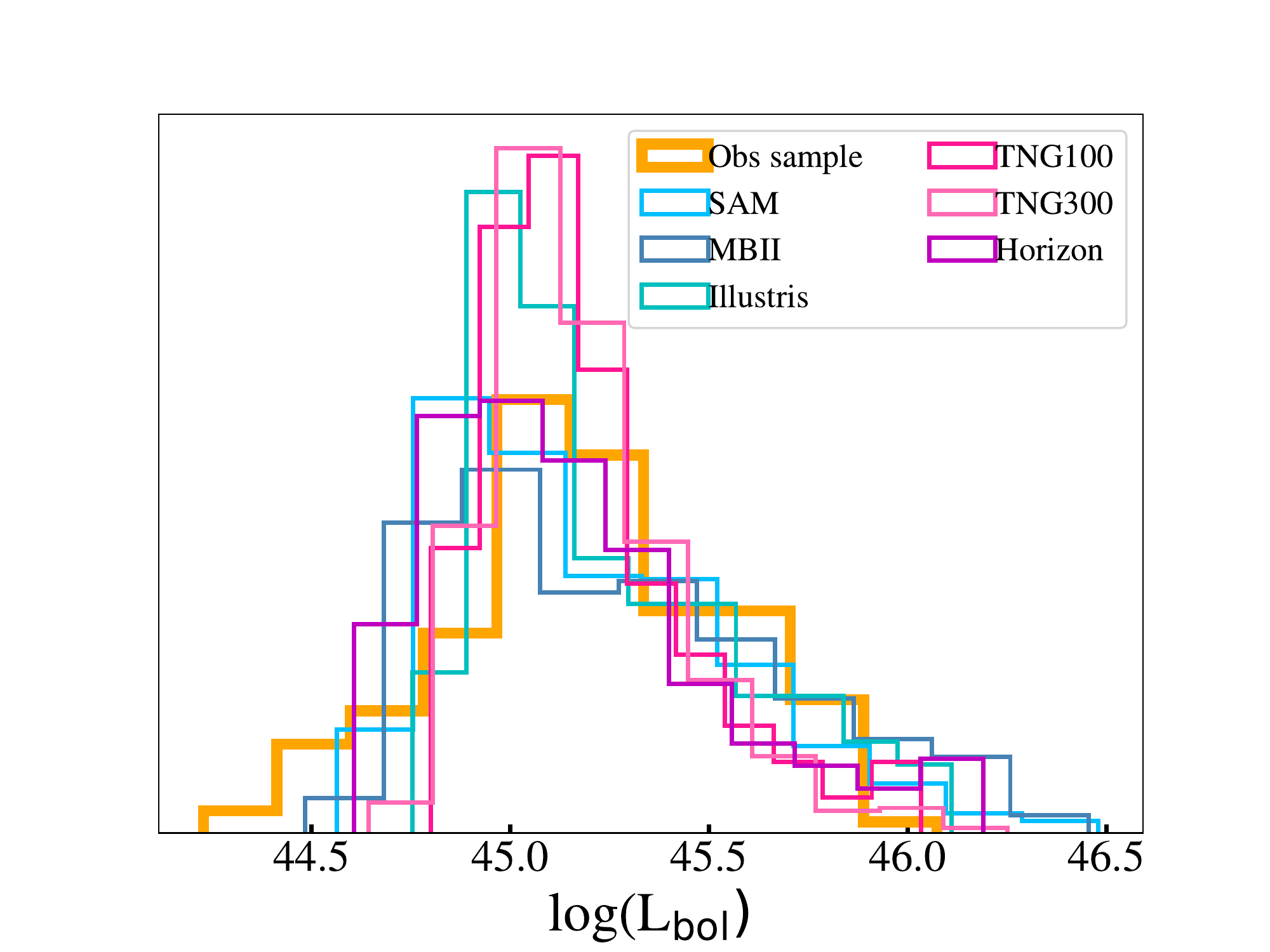}}&
\hspace*{-0.4cm} 
{\includegraphics[trim = 25mm -10mm 20mm 15mm, clip, height=0.3\textwidth]{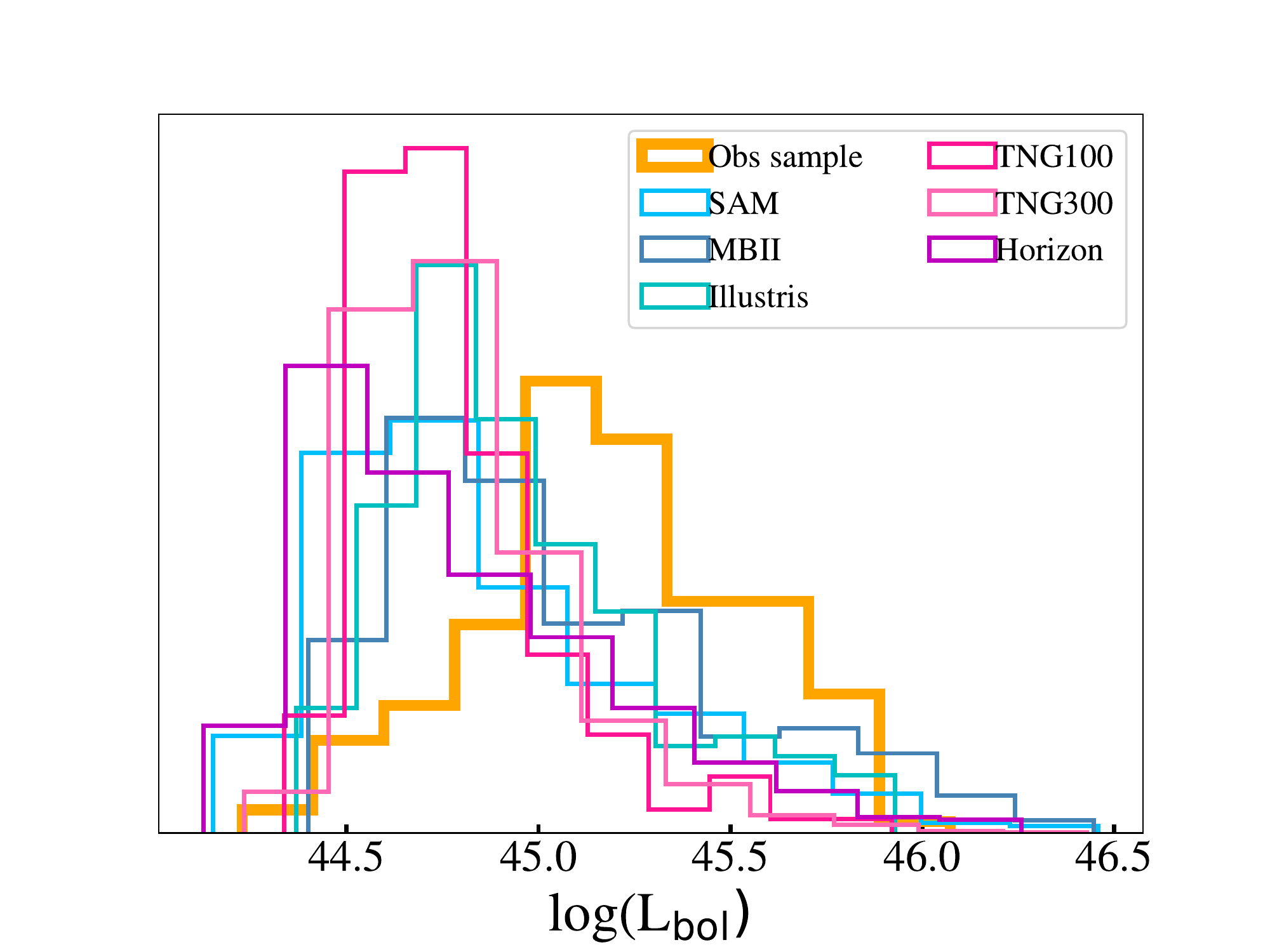}}&
\hspace*{-0.4cm} 
{\includegraphics[height=0.4\textwidth]{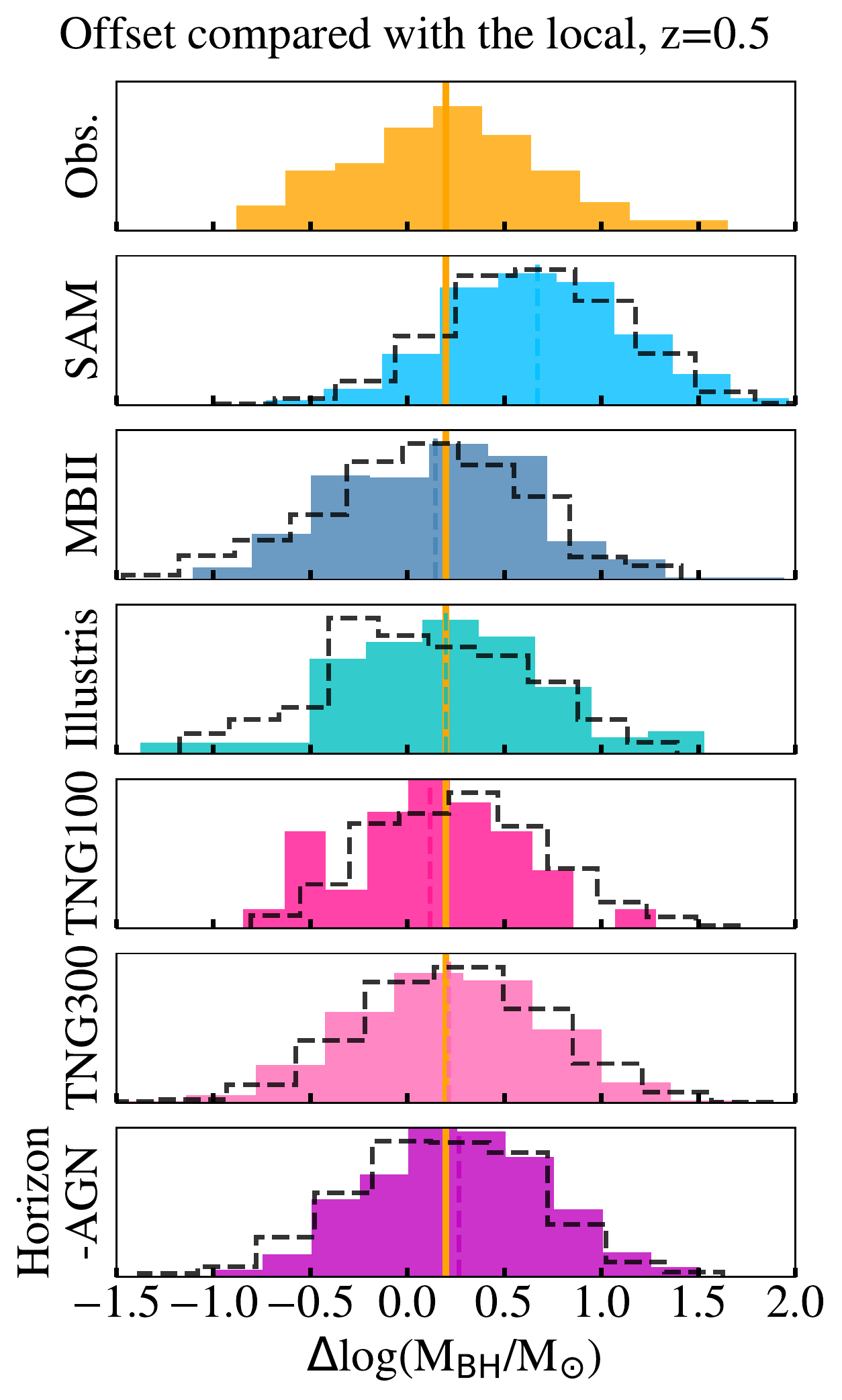}}\\
\end{tabular}
\caption{\label{fig:offsets_nochange} 
Mass comparisons using different bolometric luminosity selection. {\it Left:}~$L_{\rm bol}$ distribution of our simulated samples at $z=0.5$ tailored to match with the observations. {\it Middle:} using a different magnitude thresholds to select the simulated samples to force the distribution of the simulation to be different from the observation.  {\it Right:} offset distributions using these two selections with the dashed line being the result for a mismatched luminosity distribution. As shown, the different selections have minimal effect on the offset distributions.
}
\end{figure*}

\begin{figure}
\centering
\includegraphics[height=0.4\textwidth]{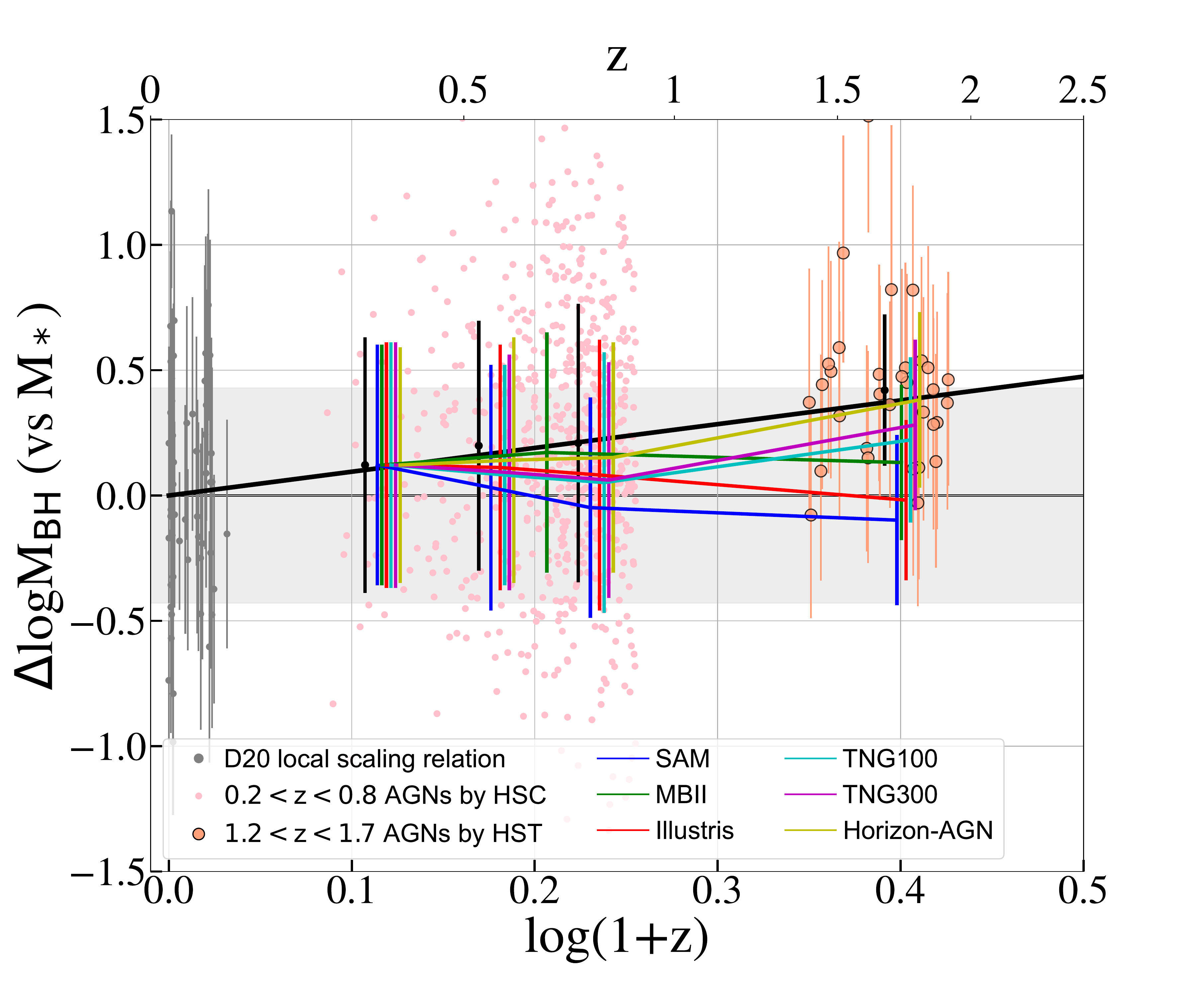}
\caption{\label{fig:offsets_vz_shiftz03} 
Same as Figure~\ref{fig:offsets_vz} but now including a shift of all simulations to match the observations at $z=0.3$.
}
\end{figure}

To avoid any possible bias raised by sample mismatch, we designed our selection (Section~\ref{subsec:add_obs_eff}) so that the distributions of $L_{\rm bol}$ and \mbh\ are similar. We also assure that the \smass\ spans similar ranges. As a test, we loosened the selection by not requiring equivalent $L_{\rm bol}$ distributions. The result shows that the offset distributions for all simulations are similar (see demonstrations in Figure~\ref{fig:offsets_nochange}) to those requiring matched $L_{\rm bol}$ distributions. This test implies that our comparison results are stable, even considering the possibility for AGN variability ($10-20\%$ level flux variation) and the lack of the obscured population being represented in our observed samples.

In the literature~\citep{Weinberger2018, Habouzit2021}, it has been noted that the mass correlations predicted by TNG300 and TNG100 are not identical. For example, TNG300 appears to have different scatter in \mbh\ at fixed \smass\ from TNG100 at \smass $>10^{10}M_{\odot}$. In addition, BH growth is more efficient in TNG100, thus causing the BH mass function to have higher normalization at the low-mass end. In our work, the differences between TNG100 and TNG300 also exist. However, the differences are milder after selection effects and random noise injection have been applied to the samples.

We considered all known observational effects, including sample selection and random noise injection, and applied these to the data products from the simulations in order to directly compare with observations. Our result shows that the scatters between the observations and simulations are very similar. Considering that the observational and simulated samples are matched to the best of our ability, their intrinsic properties are also similar. However, there may be limitations in our comparisons that may impact the results. For instance, there may be unknown observational effects that have not been properly applied to the simulations. Even so, state-of-the-art simulations likely do not capture all the physical aspects that impact the scatter. For example, the spin of BHs is not modeled in the simulation, which could change the scatter since the spin affects both the accretion rate and the energy that can be released through AGN feedback \citep{Dubois2014, 2019MNRAS.490.4133B, Habouzit2021} at some level. Even though the spin effect is still not well known and such a study is beyond the scope of this paper, a future effort may be warranted when the simulations incorporate spin.

\section{Conclusions} \label{sec:con}
We compared the observed scaling relation \mbh-\smass\ with the predictions from numerical simulations. The observational data are composed of 626 quasars at $0.2 < z < 0.8$ imaged by HSC and 32 X-ray-selected quasars at $1.2 < z < 1.7$ imaged by HST. The simulations include an SAM and five hydrodynamic simulations, i.e., MBII, Illustris, TNG100, TNG300, and Horizon-AGN. We carried out the comparisons in the observed parameter space to account for uncertainties and selection effects. To achieve this, we first injected random errors with the same observational uncertainty into the simulation and then adopted the same selection condition for the simulated data (see Figure~\ref{fig:selection}). Finally, we adopted the scaling relation from the local universe as our reference and performed comparisons using the scatter of the measurements and their central offset to the D20 local scaling relation. Our main results are summarized as follows:

\begin{enumerate}

\item{}The {\it observed} scatter predicted by the simulations is consistent with the observational measurements, i.e., $\sim0.5$~dex at $z<1$ and $\sim0.3$~dex at $z>1$ (see Figure~\ref{fig:offsets_vz} and Table~\ref{tab:sum}). This result indicates that the simulated and observed samples have consistent {\it intrinsic} scatter.

\item{}To understand how much the {\it observed} scatter is dominated by random observational error,
we rerun the estimation without injecting noise into the simulations. The obtained scatters for both $z<1$ and $z>1$ are at a similar level (i.e.,  $\sim$0.15$-$0.2 dex, see Table~\ref{tab:sum}), indicating that observational errors dominate the scatter.

\item{} Regarding the offsets of the scaling relation from the local one ($\Delta$\mbh\ at a given \smass; D20), {\it all simulations generally match the observations} with some subtle, yet notable, differences. While Illustris, MBII, and Horizon-AGN show good correspondence with observations at $z<0.6$, the comparisons at $z>0.6$ are better for SAM, MBII, TNG100, TNG300, and Horizon-AGN. Bridging the gap from \hbox{$z\sim0.7$} to $z\sim1.5$, TNG100, TNG300, and Horizon-AGN simulations match well the observed evolution of the scaling relation, i.e., the offsets are larger at higher redshift as shown in Figure~\ref{fig:offsets_vz} and Table~\ref{tab:sum}. Four out of six of the simulations have a decreasing mass offset from $z=0.2$ to $z=0.6$ while the observational mass offset increases with redshift; however, this is well below the observed scatter.
\end{enumerate}

These results are based on samples with stellar masses mainly with the range [9.5, 11.5] $M_{\odot}$. Note that the values of stellar masses in both the observations and simulations have significant uncertainty (up to a factor of two). For example, the observed \smass\ depends on the assumption of initial mass function and star formation history, while the value of \smass\ in the simulation depends on how it is defined (i.e., total mass or within 30~kpc aperture), and other subgrid models such as SN feedback and AGN feedback. In contrast, the scatter around the mean correlation is a relative quantity, which is less affected by such systematic effect. Thus, in this work we first consider the scatter as a diagnostic criterion to see whether some simulations match the data better than others. Taking (1) and (2), our results suggest that the tightness of the scaling relations has been formed since redshift 1.7, which is in contrast with the scenario of the central limit theorem~\citep{Peng2007, Jahnke2011, Hirschmann2010} that the scaling relation is a consequence of a stochastic cloud in the early universe with subsequent random mergers thereafter. In this stochastic scenario we expect the scatter of the scaling relations to increase toward higher redshift. In fact, the scatter level in the simulation without adding random noise is consistent with the {\it intrinsic} scatter estimations reported in~\citet{Ding2020, Li2021b} (i.e., $\lesssim0.35$~dex). This level is also not larger than the typical scatter of the local relations reported in the literature~\citep{Kormendy13, Gul++09, Reines2015}.

The simulations studied in this work have adopted completely different numerical techniques. Surprisingly, all of them have similar tightness of the intrinsic scaling relation and thus provide good agreement with the observations in terms of the sample dispersion. In fact, the tightness of the scaling relation likely stems from the same physics assumed in these simulations (i.e., AGN feedback). Thus, our result is {\it consistent} with the hypothesis\footnote{By stating that our comparisons are `consistent' (Section~\ref{sec:dis}), we do not mean that our hypothesis is the only plausible interpretation; there may be others.} that AGN feedback, as a causal link between supermassive BHs and their hosts, plays a key role in establishing the scaling relation.

We can gain more insight into the role of feedback by looking at the SAM model, for which multiple feedback models have been implemented. \citet{Ding2020b} compared the scaling relations obtained with the same HST sample and the SAM simulation but with a different, isotropic, AGN feedback model, and found a larger scatter ($\sim0.7$~dex) with respect to the present SAM version ($\sim0.36$~dex). We ascribed the change to the following reasons: in the new 2D model for feedback, the wave expansions stalls along the direction of the disk, and the radius where the expansion stops depends strongly on both the gas density of the disk and the AGN luminosity. 
This means that the opening angle (and hence the fraction of expelled gas) is larger when the gas density is small (because of the lower energy that has to be spent to push the gas outward) and when the AGN luminosity is large (because of the larger energy available to push the blast wave outward). These dependencies are summarized in Figure~\ref{fig:SAM}. 
Both quantities depend on the merging histories and are  related, since the AGN luminosity $L_{\rm AGN}$ depends on the available cold gas reservoir $M_{\rm gas}$.
The large efficiency of feedback in galaxies with particularly small $M_{\rm gas}$ (for given $L_{\rm AGN}$) or in those with particularly large $L_{\rm AGN}$
(for given $M_{\rm gas}$) inhibits the BH growth in all the host galaxies that are outliers with respect to the average relation between $M_{\rm gas}$ and $L_{\rm AGN}$. 
This results in a smaller scatter.

\begin{figure}
\centering
\includegraphics[trim = 0mm 0mm 0mm 30mm, clip, width=0.4\textwidth]{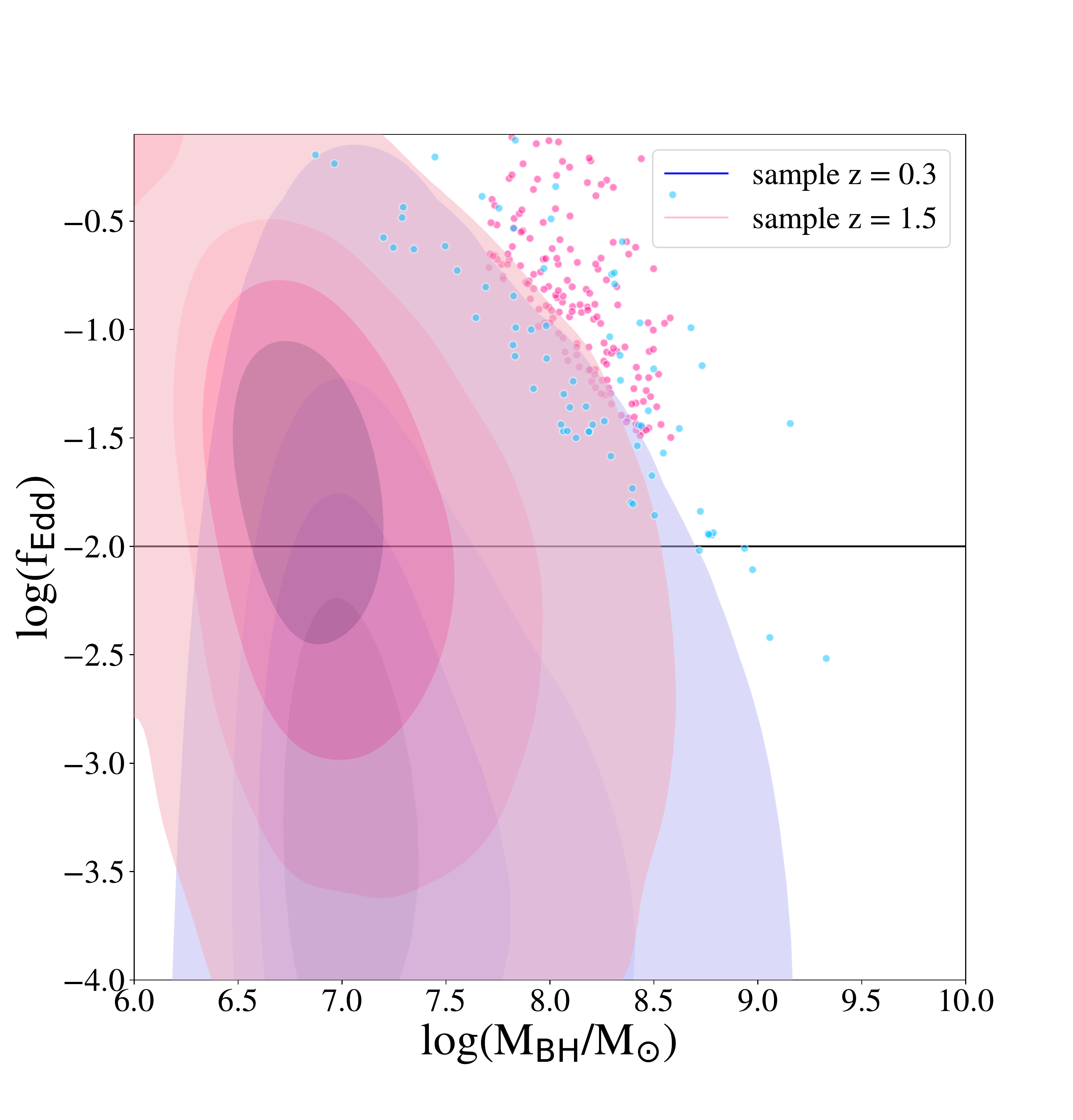}\\
\caption{\label{fig:fedd} 
Distribution of Eddington ratio and \mbh\ for the Horizon-AGN sample at redshifts $z=0.3$ and $z=1.5$. The data points indicate the distribution of the selected sample. The black line denotes the criterion in the simulation for which a supermassive BH is in a quasar (above) or radio (below) mode.}
\end{figure}

In theoretical models, AGN feedback is often assumed to consist of two distinct modes:
a quasar-heating mode where the BH accretion rates are comparable to the Eddington rate, and a radio-jet mode occurring at low accretion rates (see, e.g., Section~\ref{subsec:Horizon}). In high-redshift universe, the cold material in the early universe leads to the vigorous accretion to the supermassive BH, which drives the high accretion rates, and thus the quasar mode dominates the feedback. At low redshift, the star formation and feedback ejection reduce the cold material, leading to a lower accretion rate and a radio-mode-dominating feedback~\citep[e.g.][]{2012MNRAS.420.2662D,2016MNRAS.460.2979V,2018MNRAS.479.4056W}. Our comparison result shows that the level of intrinsic scatter in the scaling relation at redshifts up to 1.7 is consistent with low redshift (see Table~\ref{tab:sum}), which reveals the fact that at high redshift the AGN feedback described by the quasar mode may be responsible at some level for the tight correlation between supermassive BH and its host galaxy. In Figure~\ref{fig:fedd}, we demonstrate that essentially all of the quasars in Horizon-AGN that match the observed samples are at high Eddington rates. After that, the radio-jet mode starts to take control at low redshift by maintaining the tightness of the scaling relation untill the level we observe today.

Our work highlights the importance of applying measurement uncertainty and the effect of selection to the simulated data in order to make direct comparisons with observations. Such comparisons have been made in the local universe~\citep[e.g.,][]{Habouzit2021} where the measurements are relatively robust and the selection function is broad; thus, it is less crucial to ensure consistency between observations and simulations. However, beyond $z>0.2$, the scatter and the central distribution of the scaling relations are dominated by measurement uncertainty and selection effects (see Figures~\ref{fig:comparsion} and~\ref{fig:deltaMM}) and a forward modeling in the observational plane becomes essential. Indeed, those effects would hamper our understanding of whether BHs and their hosts coevolve or not. For example, from trends seen with stellar mass in Illustris and MBII (Figure~\ref{fig:deltaMM}), we found that the observations of the \mbh\ offsets as a function of \smass\  show a very different trend from the intrinsic one.

Extending this study to even higher-redshift and lower-mass galaxies (\smass\ $<10^{10}M_{\odot}$) will be very beneficial, probing closer to the epoch of formation of massive galaxies and supermassive BHs. The understanding of how and when the tight scaling relation emerged is crucial for testing theoretical models \citep{Volonteri2021}. On the observational side, the James Webb Space Telescope will provide high-quality imaging data of AGNs at redshifts up to $z\sim7$ and beyond. The upcoming measurements will represent stringent tests on the proposed physical mechanisms for the initial formation of supermassive BHs and of their subsequent evolution with galaxies.

\begin{acknowledgments}
The authors thank the anonymous referee for helpful suggestions and comments that improved this paper. We fully appreciate input from Jingjing Shi.

This work was supported by World Premier International Research Center Initiative (WPI), MEXT, Japan. Based in part on observations made with the NASA/ESA Hubble Space Telescope, obtained at the Space Telescope Science Institute, which is operated by the Association of Universities for Research in Astronomy, Inc., under NASA contract NAS 5-26555. These observations are associated with programs \#15115. Support for this work was provided by NASA through grant number HST-GO-15115 from the Space Telescope Science Institute, which is operated by AURA, Inc., under NASA contract NAS 5-26555. 
This work was supported by JSPS KAKENHI Grant Number JP22K14071.
JS is supported by JSPS KAKENHI Grant Number JP18H01251 and the World Premier International Research Center Initiative (WPI), MEXT, Japan. TT acknowledges support by the Packard Foundation through a Packard Research fellowship to TT. LB acknowledges support from NSF award AST-1909933 and Cottrell Scholar Award \#27553 from the Research Corporation for Science Advancement.
\end{acknowledgments}

\begin{deluxetable}{lcccc}
\tablecaption{Summary of the $p$-value using KS test \label{tab:pvalue}}
\tablewidth{0pt}
\tablehead{
\colhead{Simulation} &  \multicolumn{3}{c}{HSC Comparison} & \colhead{HST Comparison} \\
\cline{2-4} 
\colhead{} &  \colhead{$z\sim0.3$} & \colhead{$z\sim0.5$} & \colhead{$z\sim0.7$} & \colhead{$z\sim1.5$}
}
\startdata
SAM &  1.99e-09 & $<$1e-10  & $<$1e-10  & 7.46e-03  \\
MBII & 7.51e-01 & \multicolumn{2}{c}{3.93e-01  [$z=0.6$]}  & 2.22e-05  \\
Illustris & 5.01e-01 & 8.38e-01  & 5.98e-01  & 3.90e-05  \\
TNG100 & 1.15e-02 & 4.06e-01  & 4.84e-01  & 4.71e-01  \\
TNG300 & 2.36e-02 & 5.46e-01  & 3.10e-01  & 2.12e-01  \\
Horizon-AGN & 2.01e-01 & 4.83e-01  & 1.40e-01  & 1.95e-01  \\
\enddata
\tablecomments{These $p$-values are obtained by the KS test between the simulation and the observation based on one realization.}
\end{deluxetable}

\newcommand{\noopsort}[1]{}

\bibliographystyle{aasjournal}

\end{document}